\documentclass[showpacs,pra,floatfix,twocolumn,floatfix,superscriptaddress]{revtex4}
\pdfoutput=1
\usepackage{graphicx}
\usepackage{amssymb}
\usepackage{amsfonts}
\usepackage{amsmath}
\usepackage{epstopdf}

\newcommand {\ket}[1] {|#1 \rangle}
\newcommand {\bra}[1] {\langle#1 |}

\newcommand {\av}[1] {\langle #1 \rangle}

\begin{document}

\title{Dissipation-driven quantum phase transitions in collective spin systems}

\author{S. Morrison}
\affiliation{Institute for Theoretical Physics, University of Innsbruck, A-6020 Innsbruck, Austria}
\affiliation{Institute for Quantum Optics and Quantum Information of the Austrian Academy of Sciences, A-6020 Innsbruck, Austria}
\affiliation{Department of Physics, University of Auckland, Private Bag 92019, Auckland, New Zealand}

\author{A.~S. Parkins}
\affiliation{Department of Physics, University of Auckland, Private Bag 92019, Auckland, New Zealand}

\begin{abstract}
We consider two different collective spin systems subjected to strong dissipation -- on the same scale as interaction strengths and external fields -- and show that either continuous or discontinuous dissipative quantum phase transitions can occur as the dissipation strength is varied. First, we consider a well known model of cooperative resonance fluorescence that can exhibit a second-order quantum phase transition, and analyze the entanglement properties near the critical point. Next, we examine a dissipative version of the Lipkin-Meshkov-Glick interacting collective spin model, where we find that either first- or second-order quantum phase transitions can occur, depending only on the ratio of the interaction and external field parameters. We give detailed results and interpretation for the steady state entanglement in the vicinity of the critical point, where it reaches a maximum. For the first-order transition we find that the semiclassical steady states exhibit a region of bistability. 
\end{abstract}


\maketitle

\section{Introduction}

The field of ultracold quantum gases has recently made remarkable progress toward the implementation of (fully) tunable interacting many-body quantum systems \cite{Jaksch05}. Specifically, the degree of control in experiments allows for a precise variation of system parameters, for example interaction strengths and effective fields, such that the systems can be made to undergo transitions between different quantum phases \cite{Greiner02}.

Of particular interest are microscopic, interacting many-body systems, which have been widely studied in the context of closed systems, where quantum phase transitions (QPTs) arise due to the competition between fluctuations originating from different coherent processes in a system (e.g. tunneling versus interaction in the Bose-Hubbard Model) \cite{Osborne02,BHM98}. Although individual systems subjected to dissipation on the same scale as their characteristic frequency have been extensively studied \cite{Legget87}, the effects of dissipation on interacting many-body systems are less well-known. Recently a collective spin system with weak dissipation was studied in the context of a non-equilibrium QPT \cite{Morrison07}. It was shown that the well known second-order phase transition found in the equivalent closed system persists, with weak dissipation being responsible only for minor modifications to the system properties. However, in addition, a first-order phase transition was shown to occur exclusively due to the presence of dissipation, i.e., this phase transition is absent in the equivalent closed system case. For both types of transition, the spin-spin entanglement was shown to exhibit pronounced signatures of the criticality.

Given these non-trivial results for the case of weak dissipation, it is then naturally interesting to consider the regime of strong dissipation, i.e., dissipative rates on the same scale as the interaction strengths and external fields. In this regime marked differences are expected in comparison to the closed system case, and, specifically, new types of QPTs, driven by the dissipation, are expected to emerge \cite{Kapitulnik01,Werner05,Capriotti05,Cazalilla06,Drummond78,Carmichael80,Schneider02}. In this work we consider two models of open collective spin systems where a QPT arises solely due to a competition between fluctuations associated with Hamiltonian (coherent) dynamics and with dissipative processes. In addition to studying elementary characteristics of the phase transitions, we also study entanglement criticality and find that pronounced maxima in entanglement measures occur at the QPT. A further interesting feature that arises in the present work is that, for the second model considered, the nature of the phase transition, i.e., whether it is continuous or discontinuous, is governed by the \emph{ratio} of the spin-spin interaction strength to the effective (``magnetic'') field. This behaviour is in strong contrast to the equivalent non-dissipative models, where the character of the phase transition is governed by the nature of the interaction, i.e., by whether it is ``ferromagnetic'' or ``anti-ferromagnetic''. We also find in the latter model that within a semiclassical analysis a region of bistability arises for the first-order QPT; in the fully quantum mechanical system with finite atom number, $N$, signatures of this bistability can be identified in an atomic phase space distribution. We note that bistable behaviour and first-order non-equilibrium phase transitions have also been found in studies of optical bistability and resonance fluorescence of cooperative atomic systems \cite{Bonifacio76,Drummond78,Walls78}. However, unlike these systems, our second model involves direct spin-spin interaction terms and does not feature coherent driving of the collective atomic spin.

A brief outline of the paper is as follows. First, in Sec. \ref{sect:jx_driving_model} we briefly examine the cooperative resonance fluorescence model, which exhibits a second-order QPT as the dissipation strength is varied, and consider the steady state entanglement. Then, in Sec. \ref{sect:second_order} we focus on the dissipative Lipkin-Meshkov-Glick (LMG) model in a parameter regime where a second-order dissipation-driven QPT arises. Specifically we first present a semiclassical analysis of the phase transition and then consider the steady state entanglement behaviour across the phase transition. Next in Sec. \ref{sect:first_order} we present a similar analysis for a different parameter regime where a first-order dissipation-driven QPT occurs in the LMG model which in fact exhibits bistable behaviour in the semiclassical steady states. Finally in Sec. \ref{sect:conclusion} we will summarize our findings and give a brief outlook.

\section{Cooperative Resonance Fluorescence Model} \label{sect:jx_driving_model}

We consider here a model for cooperative resonance fluorescence as studied in \cite{Drummond78,Carmichael80,Schneider02}, which describes a collection of $N$ two-level atoms that are resonantly driven by a classical laser field and undergo collective spontaneous emission. This system can be described by the following (zero-temperature) master equation,
\begin{equation}
\dot{\rho} =  -i[\Omega J_x,\rho] + \frac{\gamma}{N} \left(2J_-\rho J_+ - J_+J_-\rho - \rho J_+J_- \right)\label{eq:jx_model_master_equation}
\end{equation}
where $\Omega$ is the strength of the coherent driving field and $\gamma$ is the collective spontaneous emission rate (i.e., $\gamma$ is proportional to the atomic density) \cite{Prefactor}. The angular momentum operators are defined in terms of the individual two level operators by $J_z = (1/2)\sum_i \sigma_z^{(i)}$, $J_\pm = \sum_i \sigma_\pm^{(i)}$, with $\sigma_\alpha^{(i)}$ the Pauli matrices for atomic spin $i$, and $J_x=(1/2)(J_++J_-)$, $J_y=(-i/2)(J_+-J_-)$. We note that this master equation possesses the exact steady state solution \cite{PuriandLawande,Drummond80},
\begin{equation}
\rho_\mathrm{ss} = \tilde{J}_-^{-1}\tilde{J}_+^{-1}, \label{rho_exact}
\end{equation}
where $\tilde{J}_\pm = J_\pm \mp i \Omega N/(2\gamma )$.

For a potential experimental realization of this system, we have in mind an ensemble of atoms coupled collectively to an optical (quantized) cavity mode and laser fields, which together drive Raman transitions between a pair of stable atomic ground states in a $\Lambda$-type configuration (similar to the setups described in \cite{Morrison07} and \cite{Dimer07}).
In particular, a pair of laser fields drive a resonant Raman transition between the atomic ground states to provide the coherent driving term in (\ref{eq:jx_model_master_equation}), while the cavity mode and another laser field drive a second, distinct Raman transition. In the (``bad cavity'') limit where the cavity field decay rate is much larger than the Raman transition rates, the cavity mode dynamics adiabatically follows the atomic dynamics and can therefore be eliminated from the model \cite{Morrison07}, yielding the dissipative term proportional to $\gamma$ in (\ref{eq:jx_model_master_equation}), with $\gamma$ the effective (cavity-mediated) collective atomic spontaneous emission rate.
Also note that in such a setup the dissipative term automatically scales with a factor of $1/N$ (in contrast to earlier studies \cite{Schneider02}), which allows the thermodynamic limit to be identified more readily and ensures that the critical point is independent of the system size $N$ in this limit.

In Sec. \ref{sect:jx_driving_model_steady_states} we first study the steady state solutions of the cooperative resonance fluorescence model. Then in Sec. \ref{sect:jx_driving_model_entanglement} we determine and analyze the steady state entanglement present in the system.

\subsection{Steady States} \label{sect:jx_driving_model_steady_states}

From the above master equation we derive the following semiclassical equations of motion for the components of the Bloch vector, $X = \av{J_x}/j$, $Y= \av{J_y}/j$, and $Z = \av{J_z}/j$ (see \cite{Morrison07} and \cite{Dimer07} for similar derivations)
\begin{subequations}
\begin{eqnarray}
\dot{X} &=& \gamma Z X,  \label{eq:jx_model_semicl(a)}\\
\dot{Y} &=& -\Omega Z + \gamma Z Y, \label{eq:jx_model_semicl(b)}\\
\dot{Z} &=&  \Omega Y - \gamma(X^2 +Y^2), \label{eq:jx_model_semicl(c)}
\end{eqnarray}
\end{subequations}
with the constraint $X^2 + Y^2 + Z^2 = 1$ corresponding to conservation of angular momentum.
For $\gamma >\gamma_{\mathrm c} \equiv \Omega$, the stable steady state solutions are given by
\begin{eqnarray}
Z_\mathrm{ss} &=& -\sqrt{1-\Omega^2/\gamma^2}, \quad X_\mathrm{ss} = 0, \quad Y_\mathrm{ss} = \Omega / \gamma. \label{eq:jx_model_semicl_ss_sols_1}
\end{eqnarray}
When $\gamma<\gamma_\mathrm{c}$ one finds that \emph{no} (semiclassical) steady state solutions exists. However, the (finite-$N$) master equation has a stable steady state solution for all $\gamma$, as given by  Eq.~(\ref{rho_exact}), which indicates that quantum fluctuations play a crucial role in determining the state of this model. This was extensively discussed in earlier works \cite{Drummond78,Carmichael80}, where an effective description for $N\gg 1$ was developed and the steady state Bloch vector components for $\gamma <\gamma_\mathrm{c}$ were shown to be
\begin{eqnarray}
Z_\mathrm{ss} &=& 0, \quad X_\mathrm{ss} = 0, \quad Y_\mathrm{ss} = \frac{\Omega}{\gamma} - \frac{\sqrt{1-(\Omega/\gamma)^2}}{\sin^{-1}(\gamma/\Omega)}. \label{eq:jx_model_semicl_ss_sols_2}
\end{eqnarray}

\begin{figure}[h!]
\centerline{\includegraphics[width=6.5cm]{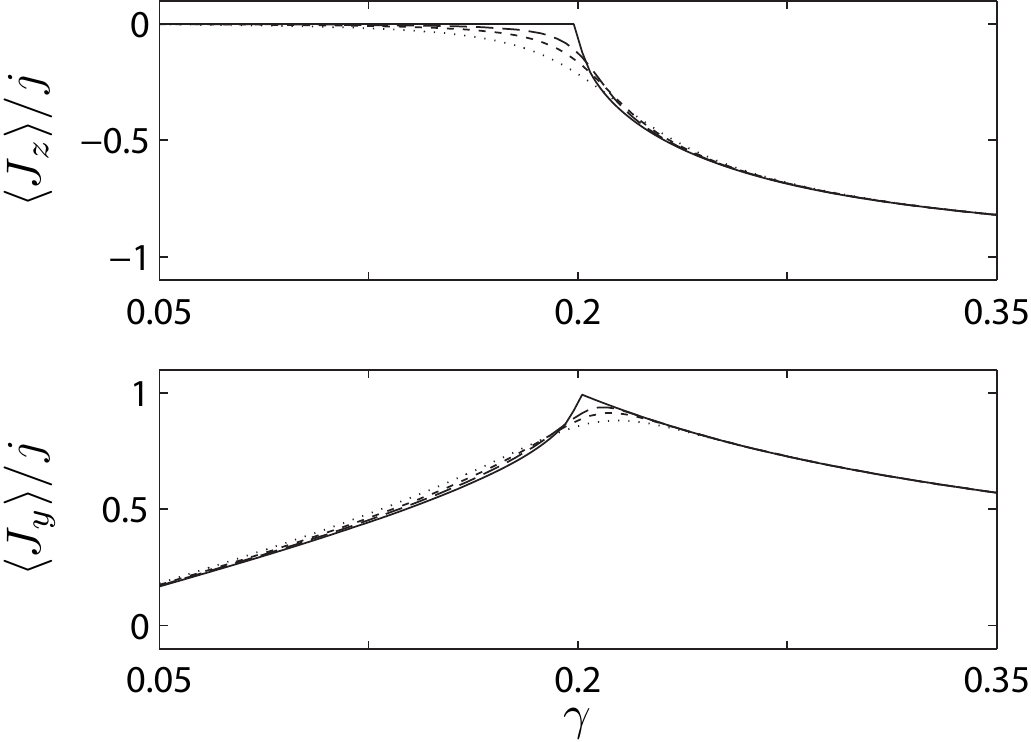}}
\caption{Semiclassical and asymptotic solutions (solid line), and finite-$N$ steady state moments for $\Omega = 0.2$, and $N=25$ (dotted line), $50$ (short dashed line), $100$ (long dashed line).} \label{fig:semicl_jx_model_first}
\end{figure}

\begin{figure}[h!]
\centerline{\includegraphics[width=6.5cm]{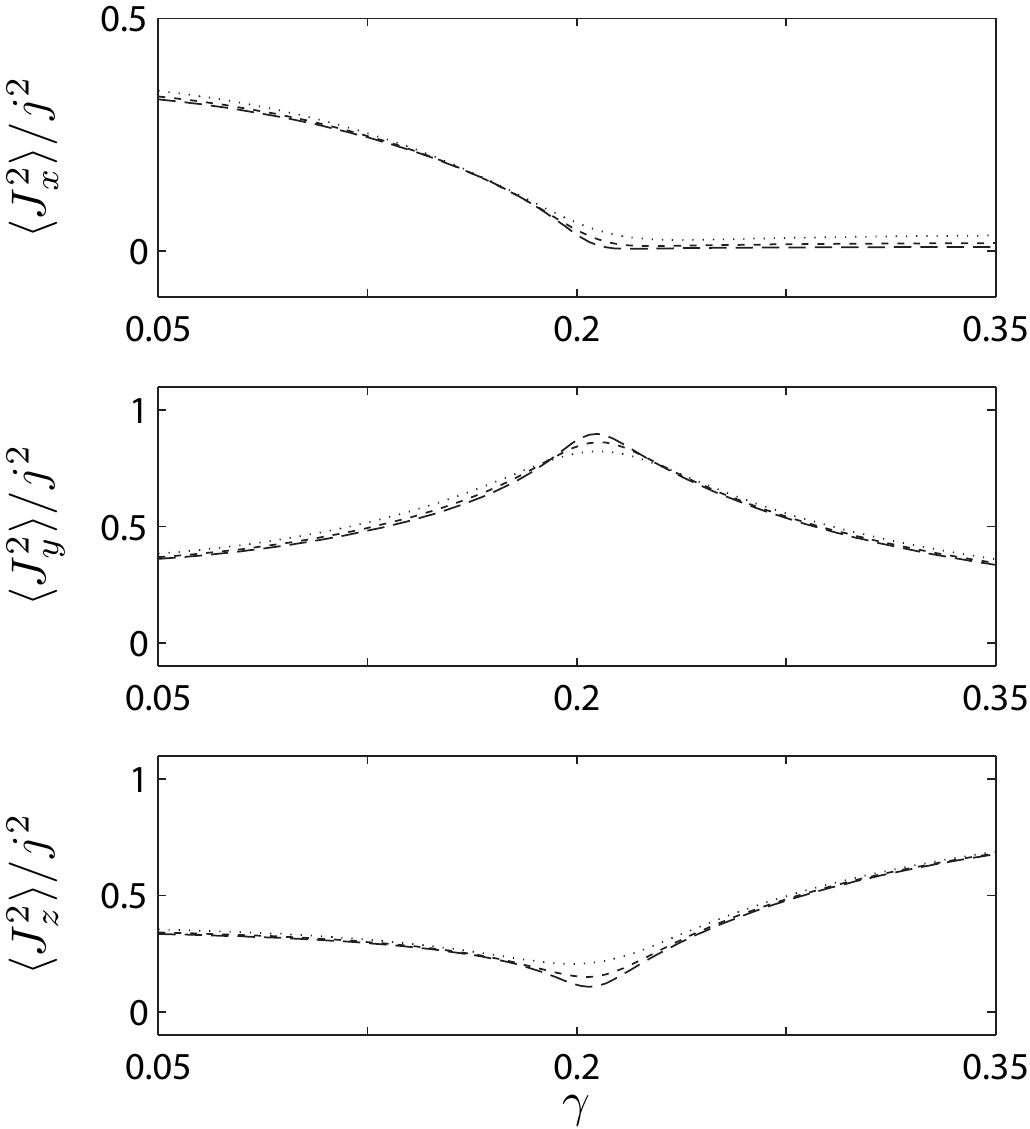}}
\caption{Finite-$N$ steady state second-order moments for $\Omega = 0.2$, and $N=25$ (dotted line), $50$ (short dashed line), $100$ (long dashed line).} \label{fig:semicl_jx_model_second}
\end{figure}

In Fig.~\ref{fig:semicl_jx_model_first} we plot the non-vanishing steady state Bloch vector components, as given by the above expressions, together with finite-$N$ solutions (computed from numerical solution of the master equation \cite{QOToolbox}) for comparison. We see that there is good agreement for sufficiently large values of $N$ (in fact, the two approaches are already in reasonable agreement for $N\simeq100$). Also, in Fig.~\ref{fig:semicl_jx_model_second} we plot finite-$N$ solutions for the steady state second-order moments, $\av{J_x^2}/j^2$, $\av{J_y^2}/j^2$, and $\av{J_z^2}/j^2$.

\subsection{Entanglement} \label{sect:jx_driving_model_entanglement}

We consider bipartite entanglement between individual atomic spins, as quantified by the rescaled concurrence, $C_\mathrm{R} = (N-1) C$, with $C$ the concurrence \cite{OrigConcurrece}, and by the phase-dependent measure max\{$0,C_\varphi$\} \cite{EntanglementCriteria}, where
\begin{equation}
C_{\varphi} \equiv 1-\frac{4}{N}\av{\Delta J_{\varphi}^2}-\frac{4}{N^2}\av{J_{\varphi}}^2 , \label{eq:finite_N_ent_criteria}
\end{equation}
with $J_\varphi = \sin(\varphi)J_x + \cos(\varphi) J_y $. Note that in Ref. \cite{Morrison07} the rescaled concurrence was found to be related to $C_\varphi$ through the relation $C_\mathrm{R} = \max_{\varphi} C_\varphi$. For our proposed realization of the model, the latter entanglement measure can in principle be determined from appropriate (quadrature variance) measurements performed on the cavity output field as explained in \cite{Morrison07}. In the present model we find that the relation $C_\mathrm{R} = \max_{\varphi} C_\varphi$ also holds, which then, indirectly, enables a measurement of the rescaled concurrence.

\begin{figure}[h!]
\centerline{\includegraphics[width=6.5cm]{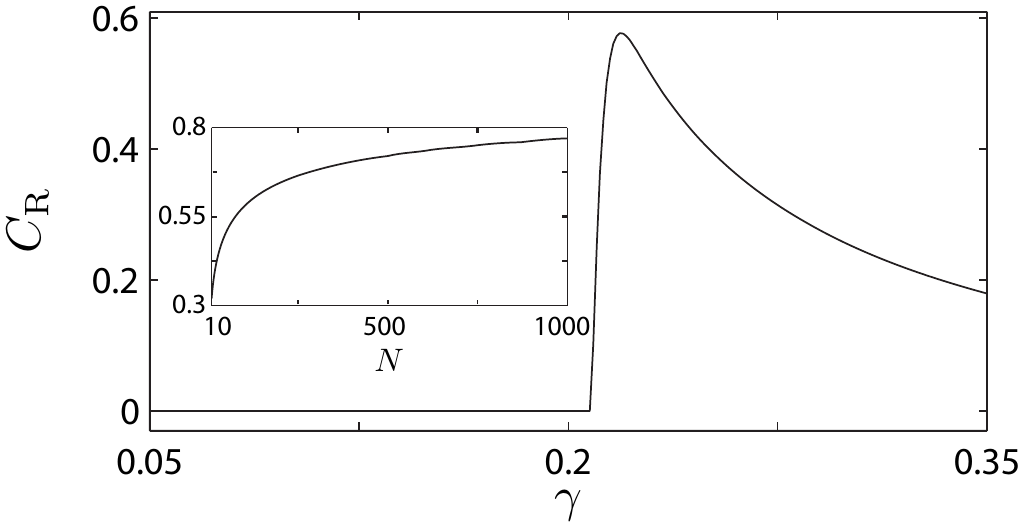}}
\caption{Rescaled concurrence $C_\mathrm{R}$ for $N=100$ and $\Omega=0.2$. Inset: Asymptotic behaviour of max($C_\mathrm{R}$) as a function of $N$ for $\Omega = 0.2$ in the vicinity of the critical point.} \label{fig:concurr_jx_model}
\end{figure}

In Fig.~\ref{fig:concurr_jx_model} we plot the rescaled concurrence as a function of the dissipation strength $\gamma$ for $N=100$ as calculated numerically from the master equation (\ref{eq:jx_model_master_equation}). We can see that the entanglement peaks close to the critical point and then very rapidly diminishes to zero below the critical point, in agreement with a previous study \cite{Schneider02}. This behaviour can be understood by considering the entanglement measure max\{$0,C_\varphi$\} as a function of the phase $\varphi$ and the dissipation strength $\gamma$. We find that above the transition, $\gamma >\gamma_\mathrm{c}$, \ $C_\varphi$ is non-zero for a broad range of $\varphi$ around $\varphi = \pi/2$. However, below the transition $
C_\varphi$ is zero for all $\varphi$ and $\gamma$ as both the mean values and fluctuations of the components of the Bloch vector, i.e., $\av{J_\alpha}^2/j^2$ and $\av{J_\alpha^2}/j^2$, scale as $j^2=N^2/4$ (see Fig.~\ref{fig:semicl_jx_model_first} and Fig.~\ref{fig:semicl_jx_model_second}). By using the exact steady state solution, given in Eq.~(\ref{rho_exact}), we are able to calculate the rescaled concurrence for large values of $N$. However, in the limit of large $\gamma$ we run into numerical difficulties and thus we will consider here only the behaviour near the critical point. In the inset of Fig.~\ref{fig:concurr_jx_model} we show the behaviour of $\mathrm{max}(C_\mathrm{R})$  as a function of $N$ in the vicinity of the critical point (since at finite $N$ the critical point depends upon $N$ it would be meaningless to consider a fixed value of $\gamma$). We see that $\mathrm{max}(C_\mathrm{R})$ continues to increase with $N$ and appears to approach the asymptotic value of $1$ (viz. the corresponding thermodynamic limit value, see below) with an approximately logarithmic scaling in $N$.

In the present context it is also useful to consider a phase space representation of the steady state as given by the spin $Q$-function,
\begin{equation}
Q_{\rm s}(\eta ) =\bra{\eta} \rho \ket{\eta},
\end{equation}
where $\ket{\eta}$ are the atomic coherent states defined by
\begin{equation}
\ket{\eta} = (1+|\eta|^2)^{-j} \sum_{m=-j}^j \sqrt{{N \choose j+m}}\eta^{j+m} \ket{j,m}_j,
\end{equation}
with  $\eta = e^{i\phi}\tan{\frac{\theta}{2}}$, where $\theta$ and $\phi$ correspond to spherical coordinates, and $\ket{j,m}$ are the Dicke states with $m\in [-j,-j+1,\ldots ,j-1,j]$ (for our system, $j=N/2$).

\begin{figure}[h!]
\centerline{\includegraphics[width=8.6cm]{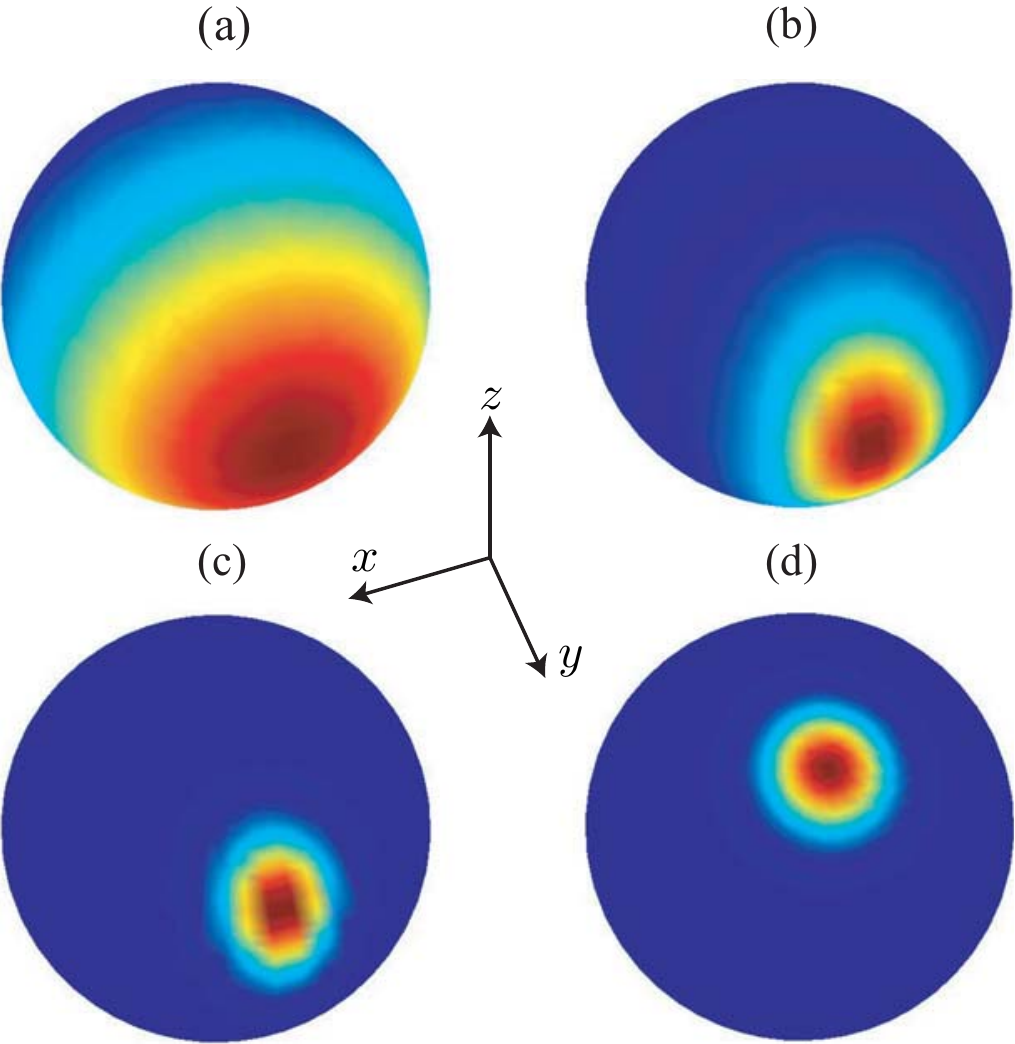}}
\caption{(Colour online) Steady state spin Q-function, $Q_\mathrm{s}(\eta)$, on the Bloch sphere for (a) $\gamma = 0.05$, (b) $\gamma = 0.15$, (c) $\gamma = 0.225$, (d) $\gamma = 0.5$, with $N=50$ and $\Omega = 0.2$. Note that dark blue corresponds to the minimum value of zero of $Q_{\rm s}(\eta)$ while dark red indicates the maximum value of $Q_{\rm s}(\eta)$.} \label{fig:spinq_jx_model}
\end{figure}

In Fig.~\ref{fig:spinq_jx_model} the spin $Q$-function, $Q_\mathrm{s}(\eta)$, is shown on the Bloch sphere for four different values of $\gamma$. We see that above the transition point $Q_\mathrm{s}(\eta)$ is a symmetric, single-peaked function centered at the corresponding semiclassical amplitude. As the critical point is approached $Q_\mathrm{s}(\eta)$ stretches and moves down to the equatorial plane. However, once below the transition point $Q_\mathrm{s}(\eta)$ remains centered at $\theta = \pi/2$ for all $\gamma$ and merely continues to spread out in size as the fluctuations increase. Eventually, as $\gamma \rightarrow 0$, $Q_\mathrm{s}(\eta)$ covers the entire Bloch sphere, in agreement with the fluctuations becoming evenly distributed in the $x$, $y$ and $z$ directions (see Fig.~\ref{fig:semicl_jx_model_second}).

Next, we consider the thermodynamic limit by linearizing the quantum fluctuations around the semiclassical steady state solutions for $N\gg1$ using the Holstein-Primakov (HP) representation \cite{Holstein40}. This basically involves replacing the spin ladder operators $J_\pm$ by bosonic creation (annihilation) operators $c_k^\dagger $ ($c_k$), where $k\in \{ <,>\}$ denotes below or above the critical point, representing the quantum fluctuations. We note that, as follows from the above discussion of the steady state representation in phase space, the linearization is only possible for $\gamma>\gamma_\mathrm{c}$, since below the critical point the state can no longer be described in terms of fluctuations centered around a semiclassical steady state solution on the Bloch sphere. However, above the critical point we can easily obtain the linearized master equation by expanding the angular momentum operators around the semiclassical steady state solution using the HP representation \cite{Holstein40}, which gives
\begin{eqnarray}
\dot{\rho}&=& \gamma_+D[c^\dagger]\rho + \gamma_- D[c]\rho + \frac{\Omega^2}{4\gamma}\left(2c \rho c  - \{ c^2 , \rho\} + \textrm{H.c.}\right), \nonumber \\
\end{eqnarray}
where $c$ and $c^\dag$ are bosonic annihilation and creation operators, respectively, $\gamma_\pm =(1-\sqrt{1\mp \frac{\Omega}{\gamma}})^2$, and we have defined the convention $D[A]\rho = 2 A \rho A^\dagger - A^\dagger A \rho - \rho A^\dagger A$.
Using this master equation, it is straightforward to compute the rescaled concurrence analytically and we find
\begin{eqnarray}
C_{\mathrm{R}} = 1-\sqrt{1-\Omega^2/\gamma^2}.
\end{eqnarray}
Away from the critical point this is in good agreement with the results shown for $N=100$ in Fig.~\ref{fig:concurr_jx_model}, whilst near the critical point the agreement between the finite-$N$ and linearized results improves with increasing system size $N$, as shown in the inset of Fig.~\ref{fig:concurr_jx_model}.

\section{Second-Order Transition in Dissipative LMG Model} \label{sect:second_order}

We now turn to the dissipative LMG model, first studied in \cite{Morrison07}, which describes a collection of $N$ interacting two-level systems in the presence of (collective) dissipation.
Specifically, as shown in \cite{Morrison07}, by considering an ensemble of atoms coupled collectively to optical cavity and coherent laser fields, with suitably tailored Raman transitions between a pair of atomic ground states, one may realize a dynamics described by the master equation
\begin{equation}
\dot{\rho} = -i[H_\mathrm{LMG},\rho] + \frac{\Gamma_a}{N} D[2J_x]\rho +
\frac{\Gamma_b}{N} D[J_+]\rho,
\label{eq:diss_lmg_master_equation}
\end{equation}
where $\Gamma_a$ and $\Gamma_b$ are tunable dissipation strengths, and the Hamiltonian is given by
\begin{equation}
H_\mathrm{LMG} = -2hJ_z - \frac{2\lambda}{N} J_x^2,
\label{eq:diss_lmg_hamiltonian}
\end{equation}
with $h$ and $\lambda$ tunable effective field and interaction strengths, respectively.

In this section we will study this model in the regime $\lambda< 2h$, where a second-order phase transition occurs. In Sec. \ref{sect:second_order_semiclassical_analysis} we analyze the steady states and non-linear dynamics of the semiclassical equations of motion. Then in Sec. \ref{sect:second_order_entanglement} we determine the steady state entanglement in the system.

\subsection{Semiclassical Analysis} \label{sect:second_order_semiclassical_analysis}

\subsubsection{Steady-state solutions}

The semiclassical equations of motion for the components of the Bloch vector, $X=\av{J_x}/j$, $Y=\av{J_y}/j$, $Z=\av{J_z}/j$, where $j=N/2$, derived from the above master equation, are given by \cite{Morrison07}
\begin{subequations}
\begin{eqnarray}
\dot{X} & = & 2h Y - \Gamma_b Z X,  \label{eq:diss_lmg_semicl(a)}\\
\dot{Y} & = & -2h X + 2\lambda Z X - \Gamma_b Z Y, \label{eq:diss_lmg_semicl(b)}\\
\dot{Z} & = & - 2\lambda X Y + \Gamma_b(X^2 +Y^2), \label{eq:diss_lmg_semicl(c)}
\end{eqnarray}
\end{subequations}
with the constraint $X^2 + Y^2 + Z^2 = 1$ corresponding to conservation of angular momentum.

The steady state solutions of these equations of motion exhibit a bifurcation at a critical dissipation strength
\begin{equation} \label{eq:lambda_c}
\Gamma_b^{\rm c} \equiv 2 \sqrt{h(\lambda - h)},
\end{equation}
provided $\lambda>h$; also note that $\Gamma_b^{\rm c}<\lambda$. For $\Gamma_b > \Gamma_b^{\rm c}$ the stable steady-state solutions are
\begin{eqnarray}
Z_\mathrm{ss} =  1 , \quad X_\mathrm{ss} =Y_\mathrm{ss} = 0, \label{eq:diss_lmg_semicl_ss_sols}
\end{eqnarray}
whilst for $\Gamma_b<\Gamma_b^{\rm c}$ they become
\begin{eqnarray}
Z_\mathrm{ss} = \frac{2h}{\Lambda}, \quad X_\mathrm{ss} = \pm \sqrt{\frac{\Lambda^2-4h^2}{2\lambda\Lambda}}, Y_\mathrm{ss} = \frac{\Gamma_b}{2h} X_\mathrm{ss} Z_\mathrm{ss}, \label{eq:diss_lmg_semicl_ss_sols_2}
\end{eqnarray}
where
\begin{equation}
\Lambda = \lambda + \sqrt{\lambda^2-\Gamma_b^2} \, .
\end{equation}

\begin{figure}[h!]
\centerline{\includegraphics[width=6.5cm]{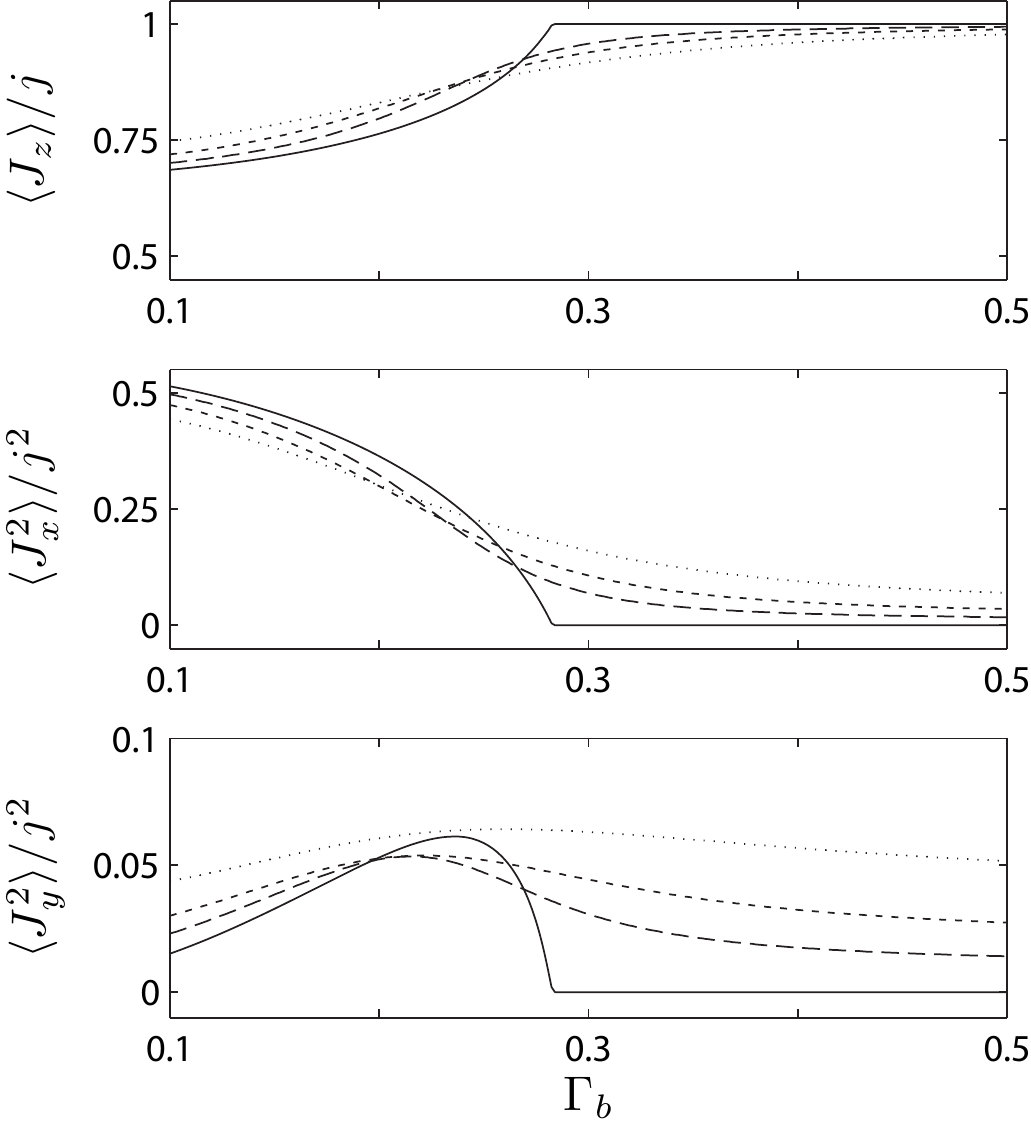}}
\caption{Semiclassical (solid line) and finite-$N$ steady state second-order moments for $h=0.2$, $\lambda=0.3$, and $\Gamma_b^\mathrm{c} = 0.28$, and $N=25$ (dotted line), $50$ (short dashed line), $100$ (long dashed line).} \label{fig:dissipative_second_order_moments}
\end{figure}

From these expressions we can see that, in the present regime of $\lambda \leq 2h$, all the semiclassical steady state solutions vary continuously across the phase transition corresponding to a second-order phase transition. In Fig.~\ref{fig:dissipative_second_order_moments} we illustrate this bifurcation, where, to facilitate a comparison between semiclassical and finite-$N$ solutions (computed from numerical solution of the master equation), we plot the second-order moments $\av{J_x^2}$ and $\av{J_y^2}$ (since the finite-$N$ master equation gives $\av{J_x}=\av{J_y}=0$ for all $\lambda$), and $\av{J_z}$. We note that the two approaches are in reasonable qualitative agreement; we expect improved quantitative agreement for increasing $N$, but unfortunately we are computationally restricted from considering much larger system sizes. This should be compared to the results of the model in the previous Sec. \ref{sect:jx_driving_model}, i.e., Fig.~\ref{fig:semicl_jx_model_first}
 ,  where the convergence between finite-$N$ and asymptotic results was much better. Note that since $\lambda \sim h$ a significant inversion of $X_{\rm ss}$ or $Z_{\rm ss}$ (as observed for the second-order transition presented in \cite{Morrison07}) does not occur here for $\Gamma_b <\Gamma_b^\mathrm{c}$.

Next let us briefly consider the detailed stability analysis of the steady state solutions, which can be readily determined by linearizing the above non-linear equations of motions (\ref{eq:diss_lmg_semicl(a)})-(\ref{eq:diss_lmg_semicl(c)}) around the steady state solutions. The resulting linearized equations of motion can be expressed as $(\dot{X},\dot{Y},\dot{Z})^T = \mathbf{M} (X,Y,Z)^T + C$, where $C$ is a (constant) three component vector, and we will consider the non-trivial eigenvalues, $\mu_\pm$, of the $3\times 3$ matrix $\mathbf{M}$ (a trivial zero eigenvalue is always present due to the constant of the motion). Above the critical point, $\Gamma_b >\Gamma_b^\mathrm{c}$, the eigenvalues of $\mathbf{M}$ are purely real and given by
\begin{equation}
\mu_\pm = -\Gamma_b \pm \Gamma_b^\mathrm{c}. \label{eq:diss_lmg_eigenvalues_trivial_solution}
\end{equation}
From this we see that the eigenvalues scale linearly with the dissipation (in contrast to the studies of our previous model \cite{Morrison07}) with $\mu_+$ going to zero at the critical point. Below the critical point, $\Gamma_b <\Gamma_b^\mathrm{c}$, the eigenvalues of $\mathbf{M}$ are given by
\begin{eqnarray}
\mu_\pm &=& -\frac{2\Gamma_b h}{\lambda+\sqrt{\lambda^2 - \Gamma_b^2}}
\pm \sqrt{2(2h^2 + 2 \Gamma_b^2-\lambda \Lambda)} \, . \label{eq:diss_lmg_eigenvalues_non_trivial_solution}
\end{eqnarray}
In the region $\Gamma_b''<\Gamma_b<\Gamma_b^\mathrm{c}$, where
\begin{equation}
\Gamma_b'' =
\sqrt{\frac{\lambda^2 - 4h^2}{2} + \frac{\lambda}{2} \sqrt{\lambda^2 + 8h^2}} \, ,
\end{equation}
the eigenvalues are also purely real, with $\mu_+$ going to zero at the critical point, whilst in the region $\Gamma_b<\Gamma_b''<\Gamma_b^\mathrm{c}$ they become complex conjugate pairs, with an imaginary part that goes to zero as $\sqrt{\Gamma_b''-\Gamma_b}$. For our characteristic parameters of $h=0.2$ and $\lambda=0.3$ we find that $\Gamma_b ''=0.25$ and $\Gamma_b^{\rm c} = 0.28$, hence $\Gamma_b''<\Gamma_b^{\rm c}$. We note from the above expression that for the case $\lambda <2h$ considered here, the eigenvalues vary smoothly across the critical point, as expected for a second-order phase transition.

\subsubsection{Time-dependent solutions}

\begin{figure}[h!]
\centerline{\includegraphics[width=8.6cm]{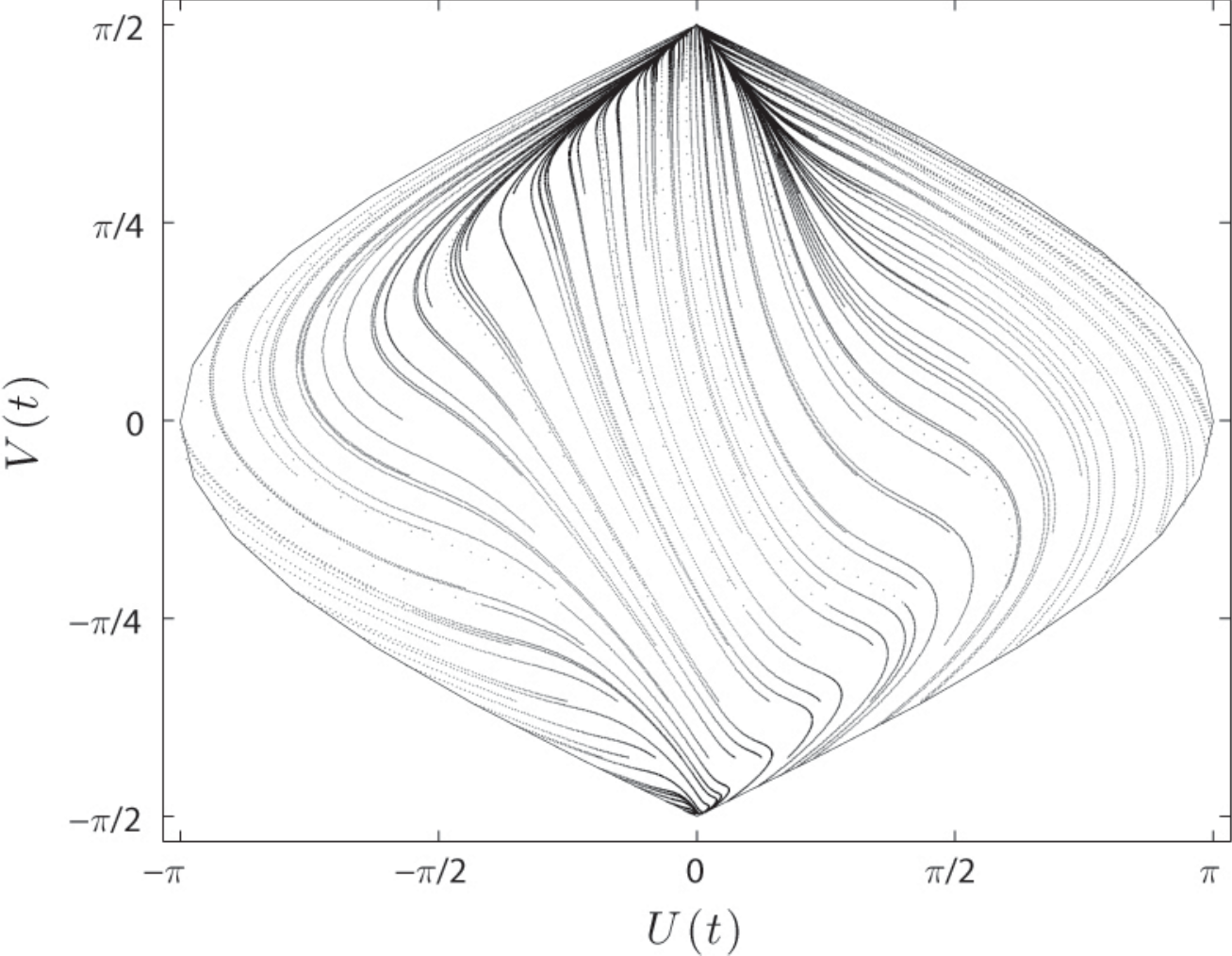}}
\caption{Trajectories for different initial conditions on the Bloch sphere using the sinusoidal projection, with $h=0.2$, $\lambda=0.3$, and $\Gamma_b=0.45$.} \label{fig:traj_triv}
\end{figure}

\begin{figure}[h!]
\centerline{\includegraphics[width=8.6cm]{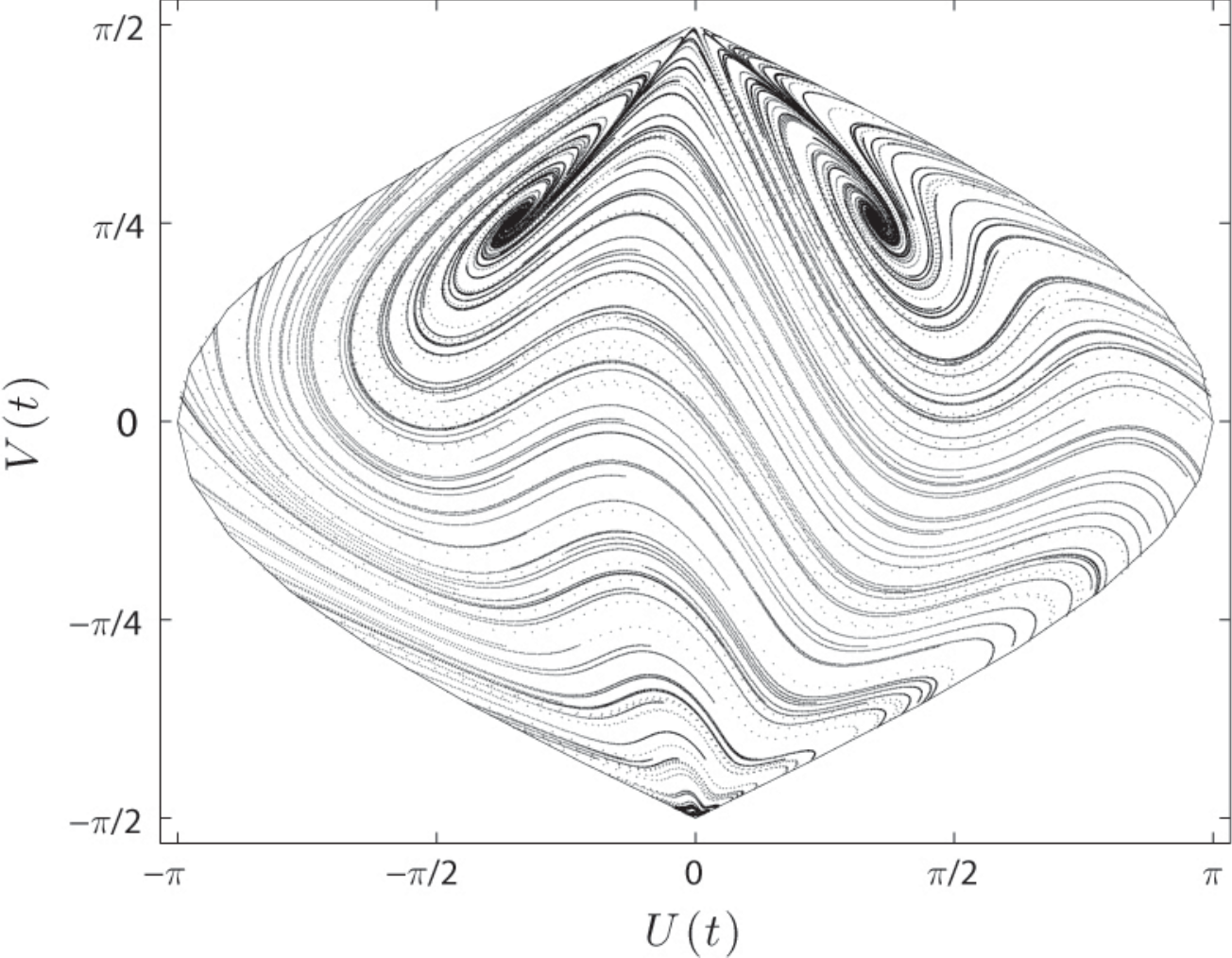}}
\caption{Trajectories for different initial conditions on the Bloch sphere using the sinusoidal projection, with $h=0.2$, $\lambda=0.3$, and $\Gamma_b=0.15$. For this choice of parameters the stable steady states are located at $U_{\rm ss} \simeq \pm 0.35\pi$, $V_{\rm ss} \simeq 0.25\pi$ (as calculated from (\ref{eq:diss_lmg_semicl_ss_sols_2})).} \label{fig:traj_non_triv}
\end{figure}

Let us now consider numerical, time-dependent solutions of the semiclassical equations of motion, i.e., of Eqs.~(\ref{eq:diss_lmg_semicl(a)})-(\ref{eq:diss_lmg_semicl(c)}). We have calculated the evolution of the Bloch components $X(t)$, $Y(t)$, and $Z(t)$ numerically for a uniform distribution of different initial states on the Bloch sphere. The resulting trajectories $\{X(t),Y(t),Z(t)\}$ are mapped from the Bloch sphere into the plane using the sinusoidal projection \cite{SinusoidalProjection}. This mapping is achieved in two steps. First, the solutions for the Bloch components are transformed into spherical polar angles $\theta(t)$ and $\phi(t)$. Next, the polar angles are transformed to new coordinates $U(t)$ and $V(t)$ via the transformation
\begin{eqnarray}
U(t) &=& (\phi(t)-\pi) \cos{(\theta(t) -\pi/2)}, \\
V(t) &=& \theta(t)-\pi/2.
\end{eqnarray}
In Fig.~\ref{fig:traj_triv} we plot the trajectories using the sinusoidal projection for a value of $\Gamma_b$ above the critical point (see the caption of Fig.~\ref{fig:traj_triv} for numerical values of parameters). We can see that all initial states terminate in the unique steady state given by Eq.~(\ref{eq:diss_lmg_semicl_ss_sols}), corresponding to the point $U_{\rm ss}=0,V_{\rm ss}=\pi/2$ in Fig.~\ref{fig:traj_triv}. Moreover we can see that all trajectories approach the stable steady state along one of two lines, corresponding to the stable steady state being a node (eigenvalues $\mu_\pm$ being purely real). Next, we consider the case where $\Gamma_b$ is below the critical value and plot the trajectories using the sinusoidal projection in Fig.~\ref{fig:traj_non_triv}. In this case we see that different initial states terminate in either one of the two stable steady states given by Eq.~(\ref{eq:diss_lmg_semicl_ss_sols_2}), the corresponding values for the points $U_{\rm ss},V_{\rm ss}$ are given in the caption of Fig.~\ref{fig:traj_non_triv}. We also note that none of the trajectories ever terminate at $U_{\rm ss} = 0,V_{\rm ss}=\pi/2$, corresponding to the steady state (\ref{eq:diss_lmg_semicl_ss_sols}), since this solution is unstable in this regime. We can also see that the trajectories form spirals centered at the stable semiclassical steady states, corresponding to the eigenvalues being complex conjugate pairs (note that $\Gamma_b<\Gamma_b''$ for the choice of parameters in Fig.~\ref{fig:traj_triv}).

\subsection{Entanglement} \label{sect:second_order_entanglement}

We consider as before the rescaled concurrence $C_\mathrm{R}$ and the phase dependent entanglement measure max\{$0,C_\varphi$\}, both numerically for finite $N$ and analytically in the thermodynamic limit. Note that in this model, for finite $N$ and also in the linearized analysis, we have $\av{J_{\varphi}}=0$ (since there are no linear driving terms in the effective Hamiltonian (\ref{eq:diss_lmg_hamiltonian}) or the corresponding linearized Hamiltonian \cite{Morrison07}), and thus $C_{\varphi} = 1-(4/N)\av{J_{\varphi}^2}$. Again, we find that $C_\mathrm{R} = \max_{\varphi} C_\varphi$, as for the above model and the model studied in \cite{Morrison07}.

\begin{figure}[h!]
\includegraphics[width=7.15cm]{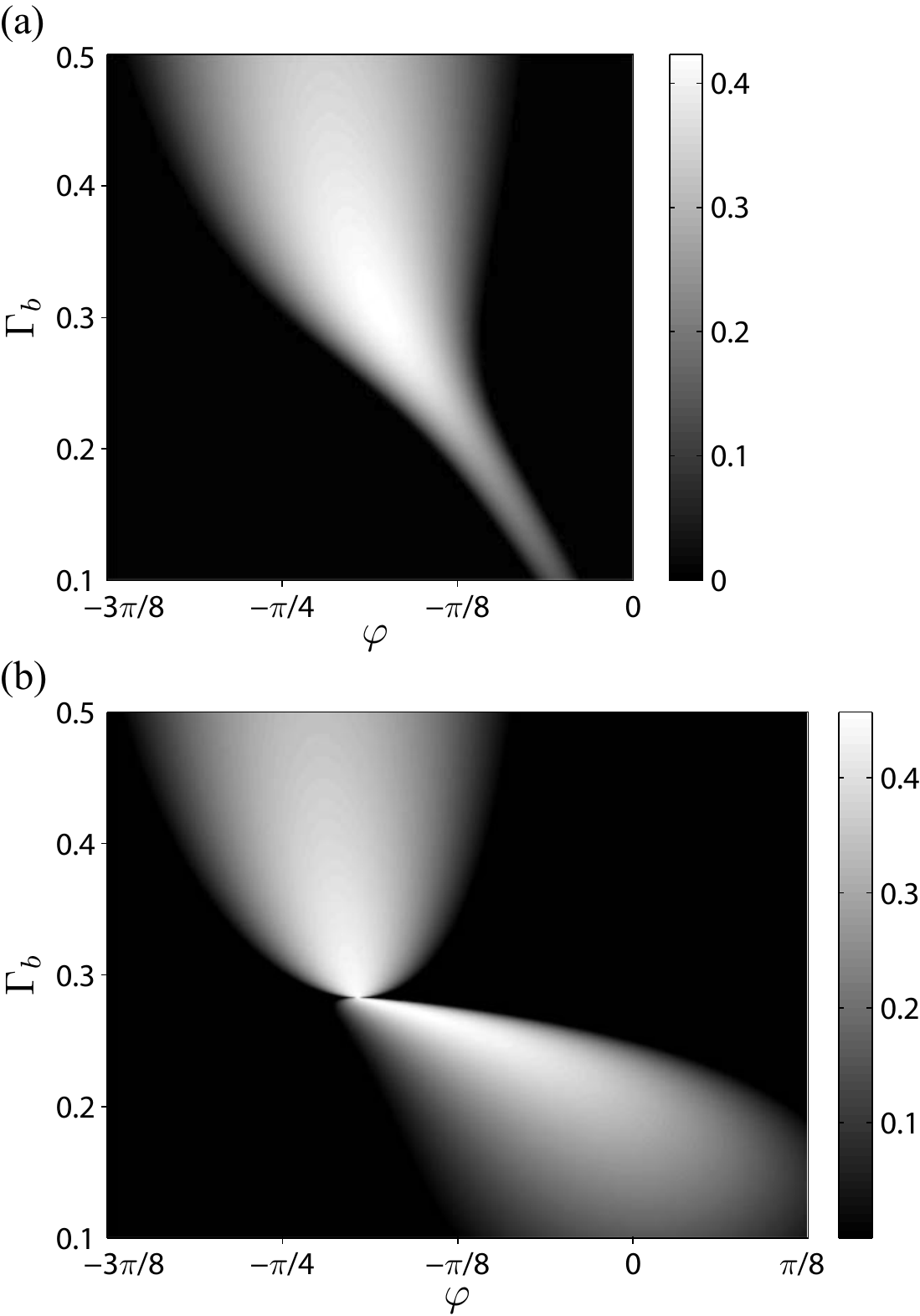}
\caption{(a) Entanglement measure $\mathrm{max}\{0,C_\varphi \}$ for $N=100$, $h=0.2$, $\lambda = 0.3$, and $\Gamma_b^\mathrm{c}=0.28$. (b) Entanglement measure $\mathrm{max}\{0,C_\varphi \}$ in the thermodynamic limit, with $h=0.2$, $\lambda = 0.3$, and $\Gamma_b^\mathrm{c}=0.28$.} \label{fig:dissipative_cvarphi_second_order}
\end{figure}

In Fig.~\ref{fig:dissipative_cvarphi_second_order} (a) we plot $\mathrm{max}\{ 0,C_\varphi\}$ as a function of $\Gamma_b$ and $\varphi$ for $N=100$ with parameters corresponding to the second-order phase transition. We see that well above the transition, $\Gamma_b > \Gamma_b^{\rm c}$, entanglement is present for a broad range of angles $\varphi$. However, as the critical point is approached the range of angles $\varphi$ which give non-zero entanglement, $C_\varphi>0$, becomes increasingly narrow. Below the transition, $\Gamma_b<\Gamma_b^{\rm c}$, the region of finite $C_\varphi$ continues to narrow and its maximum simultaneously shifts toward $\varphi = 0 $ as $\Gamma_b$ decreases.

\begin{figure}[h!]
\centerline{\includegraphics[width=8.6cm]{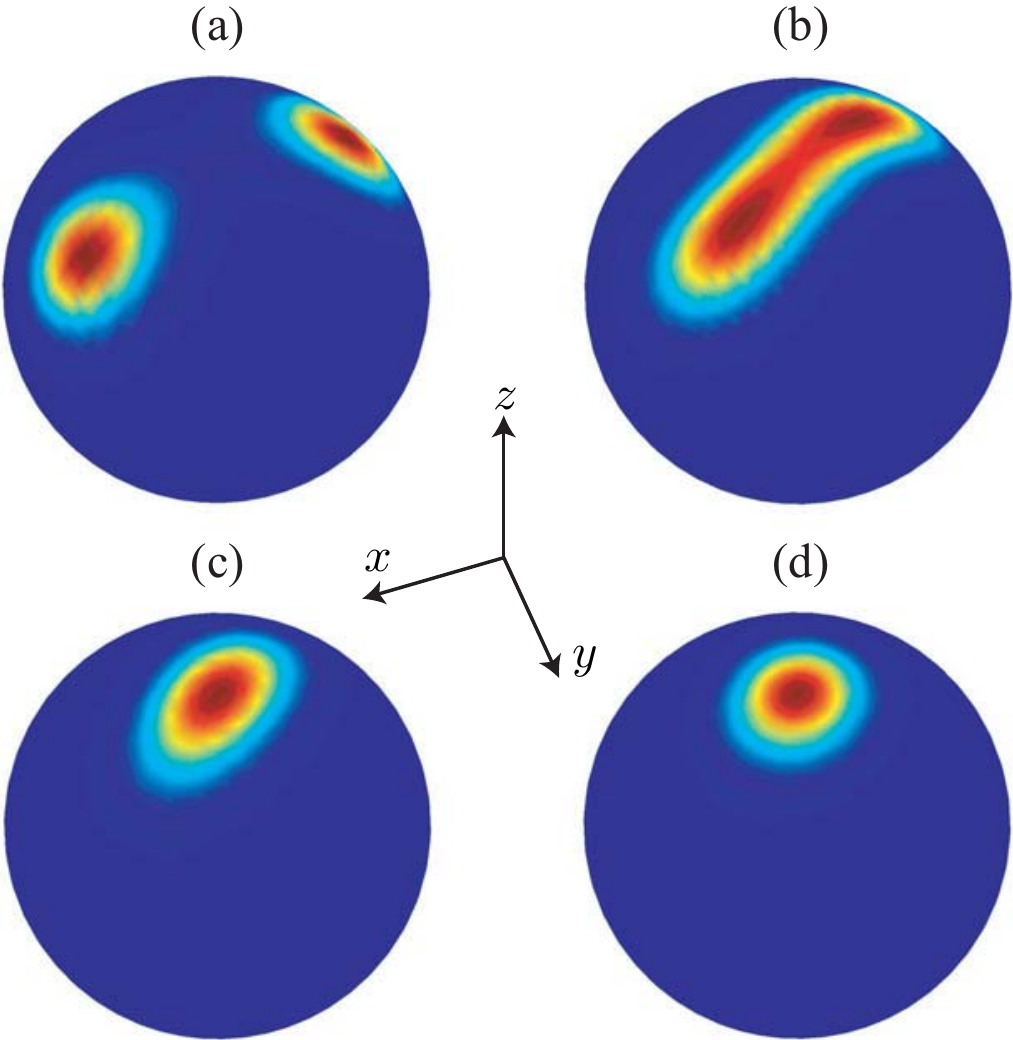}}
\caption{(Colour online) Steady state spin Q-function, $Q_\mathrm{s}(\eta)$, on the Bloch sphere for (a) $\Gamma_b = 0.1$, (b) $\Gamma_b = 0.25$, (c) $\Gamma_b = 0.4$, and (d) $\Gamma_b = 1$, with $N=50$, $h=0.2$, $\lambda = 0.3$, and $\Gamma_b^\mathrm{c}=0.28$. Note that dark blue corresponds to the minimum value of zero of $Q_{\rm s}(\eta)$ while dark red indicates the maximum value of $Q_{\rm s}(\eta)$.} \label{fig:dissipative_qspin_second_order}
\end{figure}

This behaviour can be explained by considering the spin $Q$-function. Figure \ref{fig:dissipative_qspin_second_order} displays $Q_\mathrm{s}(\eta )$ on the surface of the Bloch sphere for $N=50$ and for a series of dissipation strengths $\Gamma_b$. Above the critical point $Q_\mathrm{s}(\eta )$ is single-peaked and centered around the top of the Bloch sphere ($\theta =0$), with a significant rotation in a direction between the $x$ and $y$ axes due to the large values of dissipation. Consequently we see that although $C_\varphi$ is relatively broad above the transition, stemming from the broad shape of the lobe, its center (where it is maximal) is continually shifting toward smaller values of $\varphi$ as the rotation of $Q_\mathrm{s}(\eta)$ away from the $x$ axis decreases with decreasing dissipation.

As $\Gamma_b$ decreases towards the critical point, $Q_\mathrm{s}(\eta )$ becomes increasingly elongated along a direction between the $x$ and $y$ axes, until, at the transition, it splits into two peaks located approximately at the two semiclassical steady state amplitudes (\ref{eq:diss_lmg_semicl_ss_sols_2}). These peaks continue to move apart in phase space as the dissipation strength is decreased further, approaching the corresponding dissipation-free points at $\theta=\pi/2$ and $\phi = 0,\pi $. Correspondingly, the range of $\varphi$ over which $C_\varphi$ remains finite becomes increasingly narrow and is focussed around an axis perpendicular to that along which the two peaks lie. This narrowing of the ``width'' of $C_\varphi$ can be explained by noting that, since $\av{J_{\varphi}}=0$, we have $C_{\varphi} = 1-(4/N)\av{[\sin(\varphi)J_x + \cos(\varphi) J_y]^2}$. For decreasing dissipation strength $\Gamma_b<\Gamma_b^{\rm c}$, $\av{J
 _x^2}$ becomes of order $j^2=N^2/4$ (see Fig.~\ref{fig:dissipative_second_order_moments}), and so the optimal choice of $\varphi$ becomes more critical.

\begin{figure}[h!]
\centerline{\includegraphics[width=6.5cm]{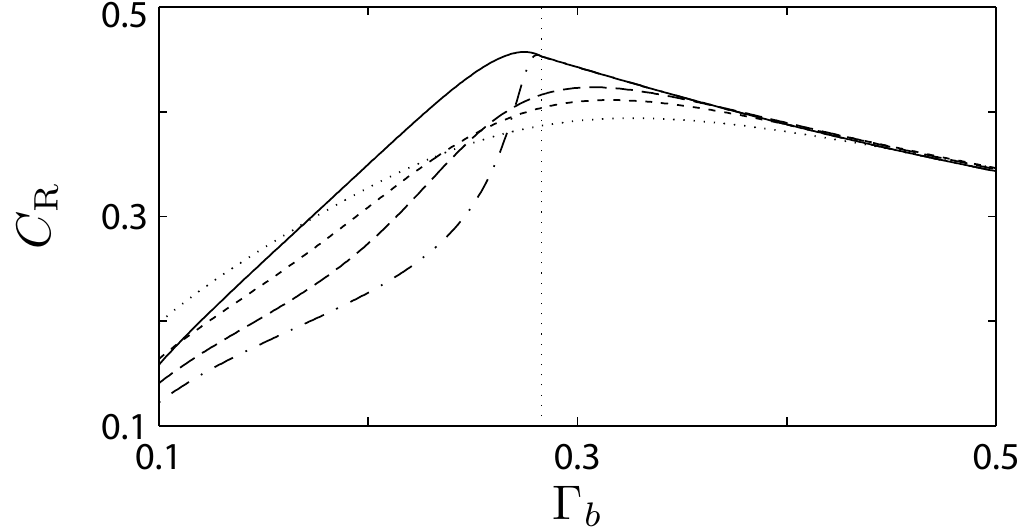}}
\caption{Rescaled concurrence $C_\mathrm{R}$ for $N=25$ (dotted line), $N=50$ (short dashed line), $N=100$ (long dashed line), the thermodynamic limit (solid line), and the thermodynamic limit in the original coordinate system (dash-dotted line, see text) with $h=0.2$, $\lambda = 0.3$, and $\Gamma_b^\mathrm{c}=0.28$. The vertical dotted line indicates the location of the critical point $\Gamma_b^{\rm c}=0.28$.} \label{fig:dissipative_concurr_second_order}
\end{figure}

In Fig.~\ref{fig:dissipative_concurr_second_order} we plot the rescaled concurrence calculated for the system sizes $N=25,50,100$ (set of dashed lines) and find that it peaks close to the critical point. In contrast to the model of the previous section, significant entanglement is present on both sides of the critical point.

We now briefly consider the thermodynamic limit where we can obtain analytic results for $N\gg1$ from the linearized HP model \cite{Morrison07} as explained in section \ref{sect:jx_driving_model_entanglement}. Since we are not directly interested in the linearized master equation we will refrain from quoting it here and refer the reader to \cite{Morrison07}.

In Fig.~\ref{fig:dissipative_cvarphi_second_order}(b) we display $C_\varphi$ in the thermodynamic limit as a function of the dissipation strength $\Gamma_b$ and the phase angle $\varphi$. We observe that above the transition, $\Gamma_b> \Gamma_b^\mathrm{c}$, the behaviour is very similar to the finite-$N$ result of Fig.~\ref{fig:dissipative_cvarphi_second_order}(a). However, the behaviour below the critical point, $\Gamma_b<\Gamma_b^\mathrm{c}$, is very different, with $C_\varphi$ non-zero for a broad range of $\varphi$ due to the linearized treatment, which only accounts for fluctuations around one of the two semiclassical steady state amplitudes (i.e., around one of the two lobes appearing in the spin $Q$-function for $\Gamma_b<\Gamma_b^{\rm c}$). Note that we can obtain plots of $\mathrm{max}\{ 0,C_\varphi\}$ similar to Fig.~\ref{fig:dissipative_cvarphi_second_order}(a) for the region $\Gamma_b < \Gamma_b^\mathrm{c}$, but determined from the linearized HP model (with a finite value of $N$), by making a rotation back to the original coordinate system and then setting, by hand, $\av{\chi_\varphi}=0$, to mimic an equal, incoherent mixture of the states associated with the two semiclassical amplitudes.

Finally, in Fig.~\ref{fig:dissipative_concurr_second_order} we also plot $C_\mathrm{R}$ as computed in the thermodynamic limit (solid line) from the linearized master equation \cite{Morrison07}. We observe that in the linearized regime the peak of the concurrence is shifted \emph{below} the critical point, whilst at the critical point there is a marked change in the behaviour. However, we can once again recover a curve within the linearized HP model that is more similar to the finite-$N$ result for $\Gamma_b < \Gamma_b^\mathrm{c}$, by following the procedure outlined at the end of the previous paragraph, which is also plotted in Fig.~\ref{fig:dissipative_concurr_second_order} (dot-dashed line).

It is interesting to note that the difference in the behaviour of max\{$0,C_\varphi$\} between the finite-$N$ and thermodynamic limit is analogous to that observed in \cite{Morrison07}. However, in \cite{Morrison07} the rescaled concurrence peaked at the critical point for both the finite-$N$ calculations and the thermodynamic limit. Thus, unfortunately, the result for max\{$0,C_\varphi$\} in the thermodynamic limit does not give us any clues to the discrepancy observed in the rescaled concurrence.

\section{First-Order Transition in Dissipative LMG Model} \label{sect:first_order}

In this section we consider the dissipative LMG model described by the master equation (\ref{eq:diss_lmg_master_equation}) in the regime $\lambda > 2h$, where we find that a discontinuous first-order transition occurs. Moreover, we find that there are in fact two transition points that encompass a region of bistability, the size of which increases with increasing $\lambda$. In Sec. \ref{sect:second_order_semiclassical_analysis} we analyze the steady states and non-linear dynamics of the semiclassical equations of motion. Then in Sec. \ref{sect:second_order_entanglement} we determine the steady state entanglement in the system.

\subsection{Semiclassical Analysis} \label{sect:first_order_semicl}

\subsubsection{Steady state solutions}

We consider, as previously, the semiclassical equations of motion given by (\ref{eq:diss_lmg_semicl(a)})-(\ref{eq:diss_lmg_semicl(c)}) and begin by studying their steady state solutions. In the region $\Gamma_b>\lambda$ the stable steady state solutions are given by (\ref{eq:diss_lmg_semicl_ss_sols}), whilst for $\Gamma_b<\Gamma_b^{\rm c}$ they are given by (\ref{eq:diss_lmg_semicl_ss_sols_2}). However, in the region $\Gamma_b^{\rm c}<\Gamma_b<\lambda $ \emph{both} steady state solutions (\ref{eq:diss_lmg_semicl_ss_sols}) and (\ref{eq:diss_lmg_semicl_ss_sols_2}) are in fact stable, i.e., the semiclassical system is bistable \cite{Bistability}. These two solutions each exhibit a discontinuity, associated with a first-order phase transition, at the critical points $\Gamma_b^{\rm c}$ and $\lambda$, respectively. Note that for values of $\lambda$ not significantly larger than $2h$, the bistable region i
 s in fact very small, with $\Gamma_b^{\rm c}\simeq \lambda$, and consequently the distinction between steady states in this region becomes somewhat redundant \cite{Morrison07}. However, in the regime $\lambda \gg 2h$ the extent of the bistable region becomes significant and both stable steady states must be considered. In fact, in this situation we can expect that a complete description in terms of a single steady state will not be possible. We will focus on the regime of large $\lambda$, and study the system primarily for the characteristic parameters $\{h=0.2,~\lambda=0.75,~\Gamma_a=0.01\}$.

\begin{figure}[h!]
\centerline{\includegraphics[width=6.5cm]{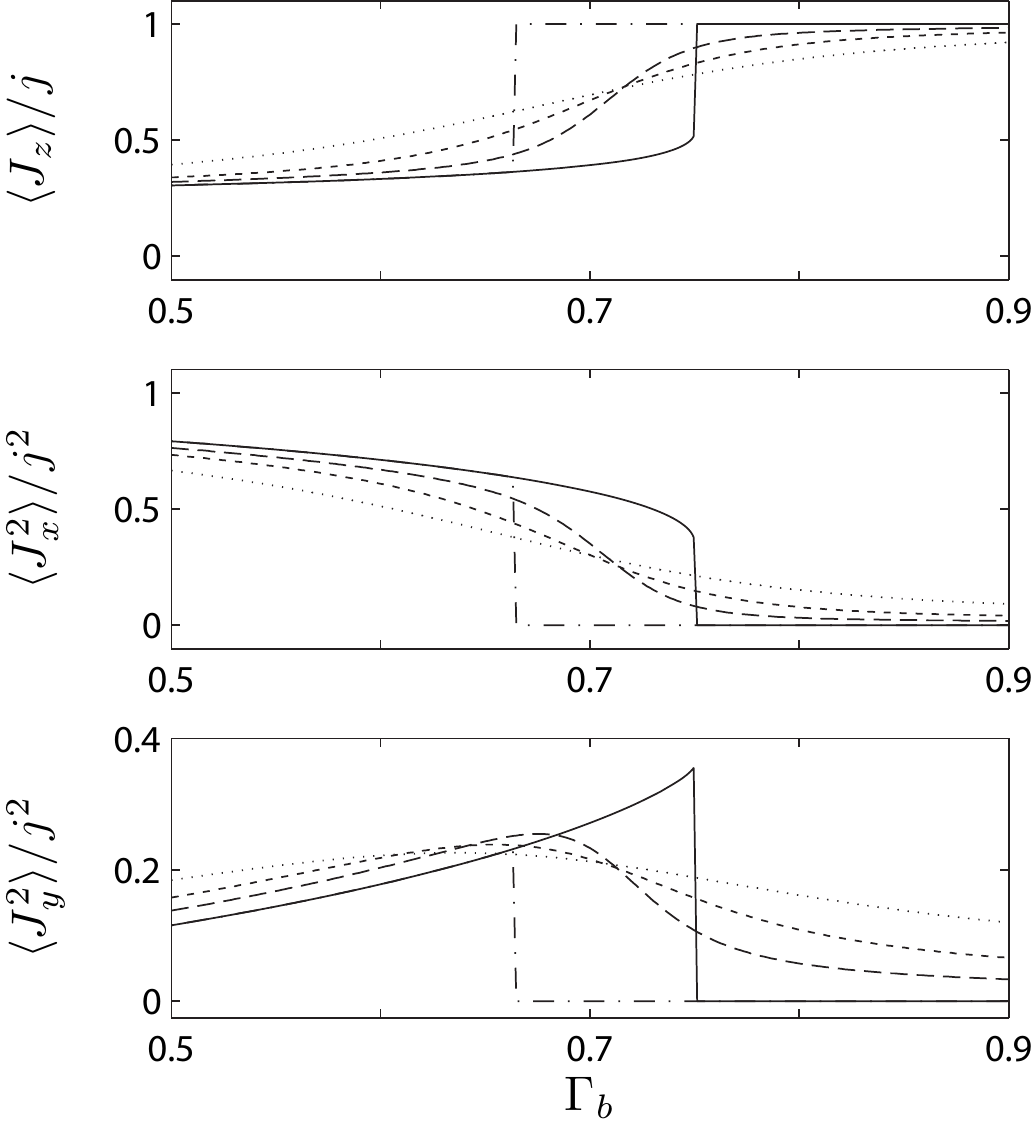}}
\caption{Stable semiclassical solutions, (\ref{eq:diss_lmg_semicl_ss_sols_2}) (solid line), and (\ref{eq:diss_lmg_semicl_ss_sols}) (dash-dotted line), with $h=0.2$, $\lambda=0.75$, and $\Gamma_b^\mathrm{c}=0.66$. Also plotted are the finite-$N$ steady state moments for $N=25$ (dotted line), $50$ (short dashed line), $100$ (long dashed line).} \label{fig:dissipative_first_order_moments}
\end{figure}

The behaviour of the semiclassical steady state solutions as a function of $\Gamma_b$ is illustrated in Fig.~\ref{fig:dissipative_first_order_moments}, together with finite-$N$ solutions. Outside the bistable region, convergence of the finite-$N$ solutions towards the semiclassical results is evident, while inside the bistable region the finite-$N$ solutions appear to approach the semiclassical branch corresponding to Eq.~(\ref{eq:diss_lmg_semicl_ss_sols_2}), although the rate of convergence with increasing $N$ is clearly much slower, and indeed the finite-$N$ curves are suggestive of some degree of ``averaging'' between the distinct semiclassical steady states. (In fact, consideration of the spin $Q$-function later in this section will support this picture.)

At the critical points, the semiclassical moments ``jump'' by a larger amount as $\lambda$ is increased; specifically the size of the jump at the critical points is quantified by $\Delta \equiv 1 - Z_\mathrm{ss}$, where $Z_\mathrm{ss}$ is given by (\ref{eq:diss_lmg_semicl_ss_sols_2}). For the critical point $\Gamma_b^{\rm c}$ this is given by $\Delta = 1 - h/(\lambda-h)$ and for the critical point $\lambda$ by $\Delta = 1-2h/\lambda$, hence we see that at either critical point the largest possible jump occurs in the limit $\lambda\gg h$.

\begin{figure}[h!]
\centerline{\includegraphics[width=6.5cm]{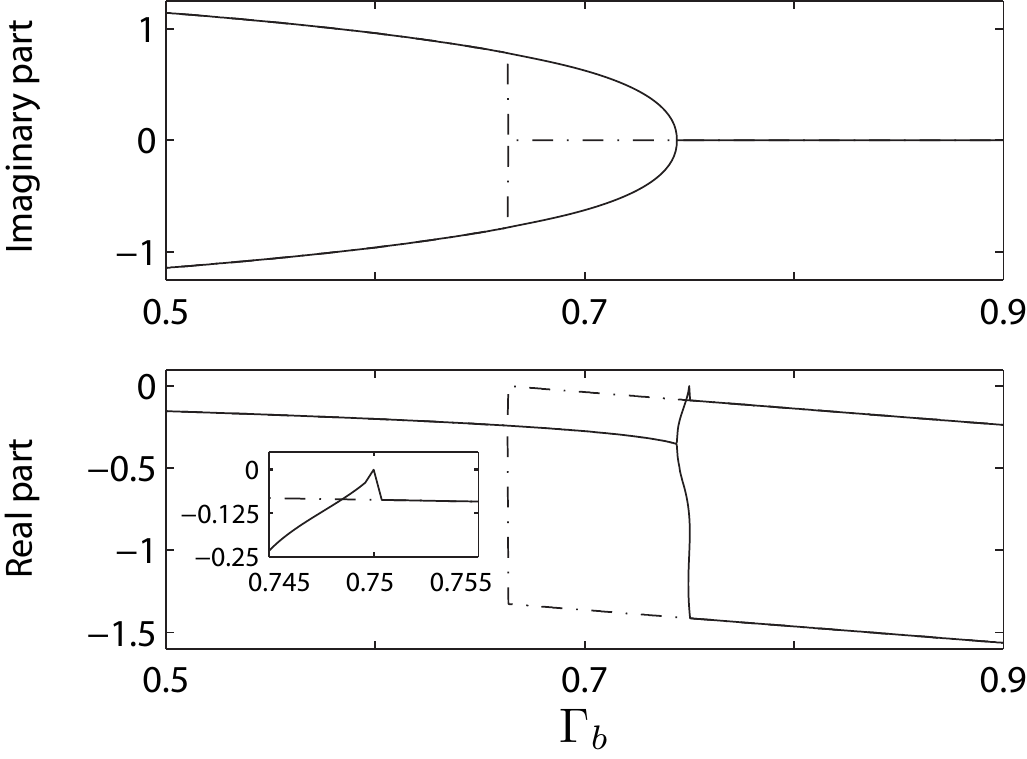}}
\caption{Eigenvalues  of the linearized equations of motion, $\mu_\pm$, as given by (\ref{eq:diss_lmg_eigenvalues_non_trivial_solution}) (solid line), and by (\ref{eq:diss_lmg_eigenvalues_trivial_solution}) (dash-dotted line), for $h=0.2$, $\lambda = 0.75$, and $\Gamma_b^\mathrm{c}=0.66$. The inset shows a magnification of the real part of $\mu_\pm$ near the critical point at $\Gamma_b=\lambda$.} \label{fig:eig_lin_first}
\end{figure}

Now let us again consider the stability of the steady state solutions by examining the non-trivial eigenvalues $\mu_\pm$ of the matrix $\mathbf{M}$ describing the linearized semiclassical equations of motion. For the steady state solution (\ref{eq:diss_lmg_semicl_ss_sols}), stable in the region $\Gamma_b^{\rm c}<\Gamma_b$, the eigenvalues are again given by (\ref{eq:diss_lmg_eigenvalues_trivial_solution}), whilst for the steady state solutions (\ref{eq:diss_lmg_semicl_ss_sols}), stable in the region $\Gamma_b<\lambda$, they are given by (\ref{eq:diss_lmg_eigenvalues_non_trivial_solution}). We plot these eigenvalues in Fig.~\ref{fig:eig_lin_first} and note the discontinuities in  (\ref{eq:diss_lmg_eigenvalues_trivial_solution}) and (\ref{eq:diss_lmg_eigenvalues_non_trivial_solution})  at $\Gamma_b^{\rm c}$ and $\lambda$, respectively, as expected for a first-order transition.
These discontinuities coincide with $\mu_+$, as given by (\ref{eq:diss_lmg_eigenvalues_trivial_solution}), going to zero at $\Gamma_b^{\rm c}$, and $\mu_+$, as given by (\ref{eq:diss_lmg_eigenvalues_non_trivial_solution}), going to zero at $\lambda$.

In the bistable region the behaviour of the eigenvalues associated with the steady states (\ref{eq:diss_lmg_semicl_ss_sols_2}) is quite similar to that of the eigenvalues for $\Gamma_b <\Gamma_b^{\rm c}$ for the second-order transition of Sec. \ref{sect:second_order}. Note, however, that whilst for $\lambda < \frac{h}{2}(3+\sqrt{5})$ (the case encountered previously in Sec. \ref{sect:second_order}) one finds $\Gamma_b''<\Gamma_b^{\rm c}$, if $\lambda>\frac{h}{2}(3+\sqrt{5})$ (corresponding to the case plotted in Fig.~\ref{fig:eig_lin_first}) we find that $\lambda>\Gamma_b''>\Gamma_b^\mathrm{c}$ (this was already noted in \cite{Morrison07}).

Finally, over essentially all of the bistable region $\Gamma_b^{\rm c} < \Gamma_b <\lambda$ we observe that the real part(s) of the eigenvalues associated with the solutions (\ref{eq:diss_lmg_semicl_ss_sols_2}) are smaller (i.e., more negative) than the largest real part of the eigenvalues associated with the solutions (\ref{eq:diss_lmg_semicl_ss_sols}). This points to the steady state solutions (\ref{eq:diss_lmg_semicl_ss_sols_2}) being more stable than the solutions (\ref{eq:diss_lmg_semicl_ss_sols}) over the majority of the bistable region, which is consistent with the apparent convergence of the finite-$N$ results with increasing $N$ towards the solutions (\ref{eq:diss_lmg_semicl_ss_sols_2}). We note, however, that for significantly larger values of $\lambda$, where the bistable region is much larger, either steady state solution can be more stable dependin
 g on the choice of $\Gamma_b$ within the bistable region, and thus a convergence of the finite-$N$ results toward one of the two semiclassical solutions is not observed.

\subsubsection{Time-dependent solutions}

\begin{figure}[h!]
\centerline{\includegraphics[width=8.6cm]{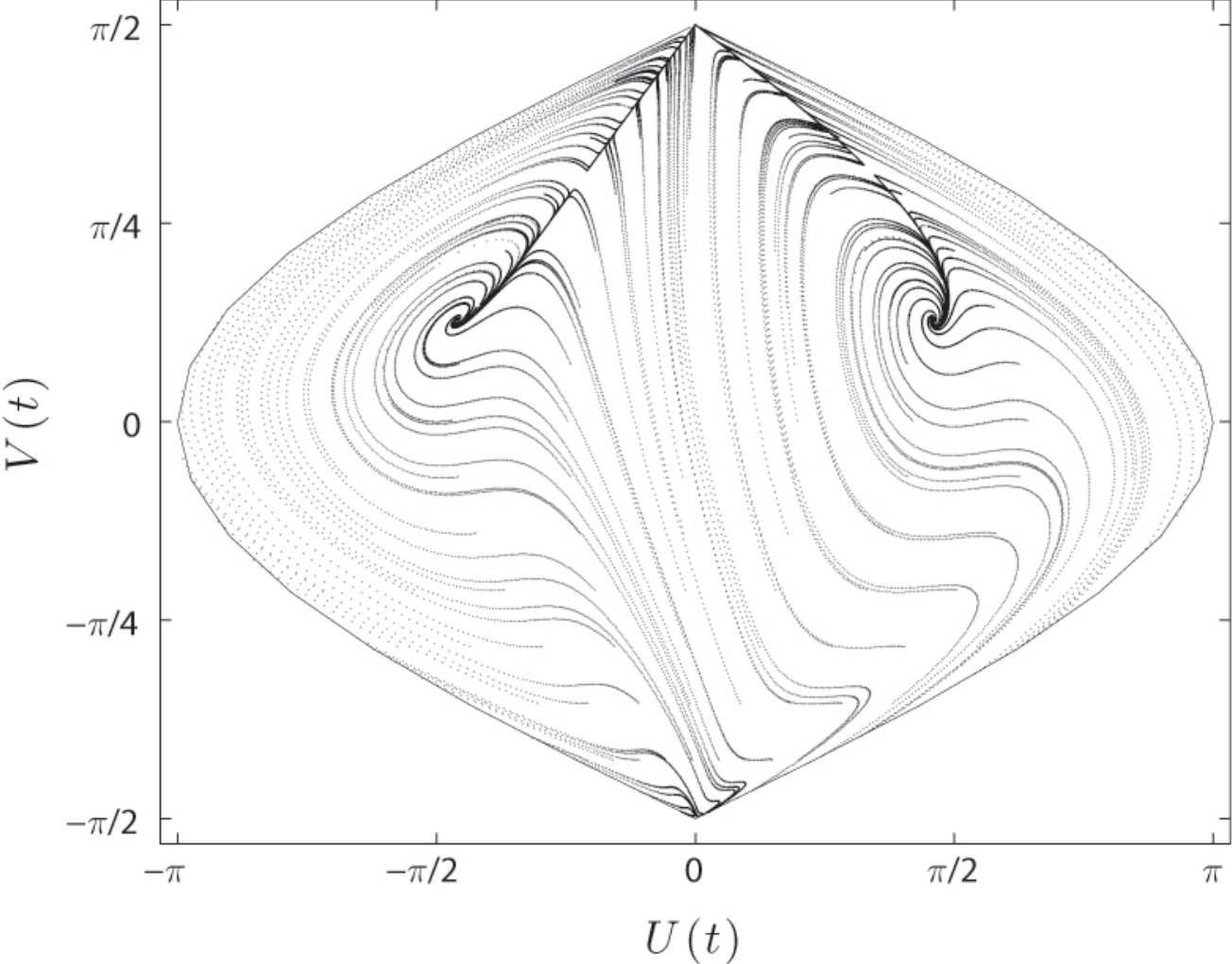}}
\caption{Trajectories on the Bloch sphere for different initial conditions, using the sinusoidal projection. Parameters are $h=0.2$, $\lambda=0.75$, and $\Gamma_b=0.7$. For this choice of parameters the broken phase stable steady states are located at $U_{\rm ss} \simeq \pm 0.56\pi$, $V_{\rm ss} \simeq 0.13\pi$ (as calculated from (\ref{eq:diss_lmg_semicl_ss_sols_2})), whilst the normal phase steady state solution is located at $U_{\rm ss}=0$, $V_{\rm ss} =\pi/2$. Note that the unstable solution described in the text is located at $U_{\rm ss} \simeq \pm 0.21\pi$, $V_{\rm ss} \simeq 0.31\pi$ for the parameters considered.} \label{fig:traj_tristable}
\end{figure}

Time dependent solutions of the semiclassical equations of motion are now examined, again using the sinusoidal projection defined earlier. In particular, in Fig.~\ref{fig:traj_tristable} we plot trajectories $\{ U(t),V(t)\}$ for a value of $\Gamma_b$ within the bistable region $\Gamma_b^{\rm c} < \Gamma_b <\lambda$ \cite{Trajectories}. We see that, in this case, different initial states evolve to \emph{either} the steady state (\ref{eq:diss_lmg_semicl_ss_sols}), (corresponding to $U_{\rm ss}=0,V_{\rm ss}=\pi/2$), \emph{or} to either of the two steady states (\ref{eq:diss_lmg_semicl_ss_sols_2}) (corresponding to $U_{\rm ss}\simeq\pm 0.56\pi,V_{\rm ss}\simeq 0.13\pi$). Again we see that the way in which the trajectories approach the respective steady states is directly related to the type of eigenvalue.

In Fig.~\ref{fig:traj_tristable} we note two distinct ``gaps''  centered at the points $\{ U \simeq \pm 0.21\pi , V \simeq 0.31\pi\}$. These points in fact correspond to {\em unstable} steady state solutions of the semiclassical equations of motion, given by
\begin{eqnarray}
Z_\mathrm{ss}^{\rm u} = \frac{2h}{\Lambda'}, ~X_\mathrm{ss}^{\rm u} = \pm \sqrt{\frac{\Lambda'^2-4h^2}{2\lambda\Lambda'}},~Y_\mathrm{ss}^{\rm u} = \frac{\Gamma_b}{2h} X_\mathrm{ss}^{\rm u} Z_\mathrm{ss}^{\rm u} ,
\end{eqnarray}
where $\Lambda' = \lambda - \sqrt{\lambda^2-\Gamma_b^2}$. These points separate trajectories that terminate in different (stable) steady states, i.e., trajectories that pass ``above'' (``below'') these points terminate in the steady state $\{ U_{\rm ss}=0,V_{\rm ss}=\pi/2\}$ ($\{ U_{\rm ss}=\pm 0.56\pi,V_{\rm ss}\simeq 0.13\pi\}$).

\subsection{Entanglement} \label{sect:first_order_entanglement}

\begin{figure}[h!]
\includegraphics[width=8.6cm]{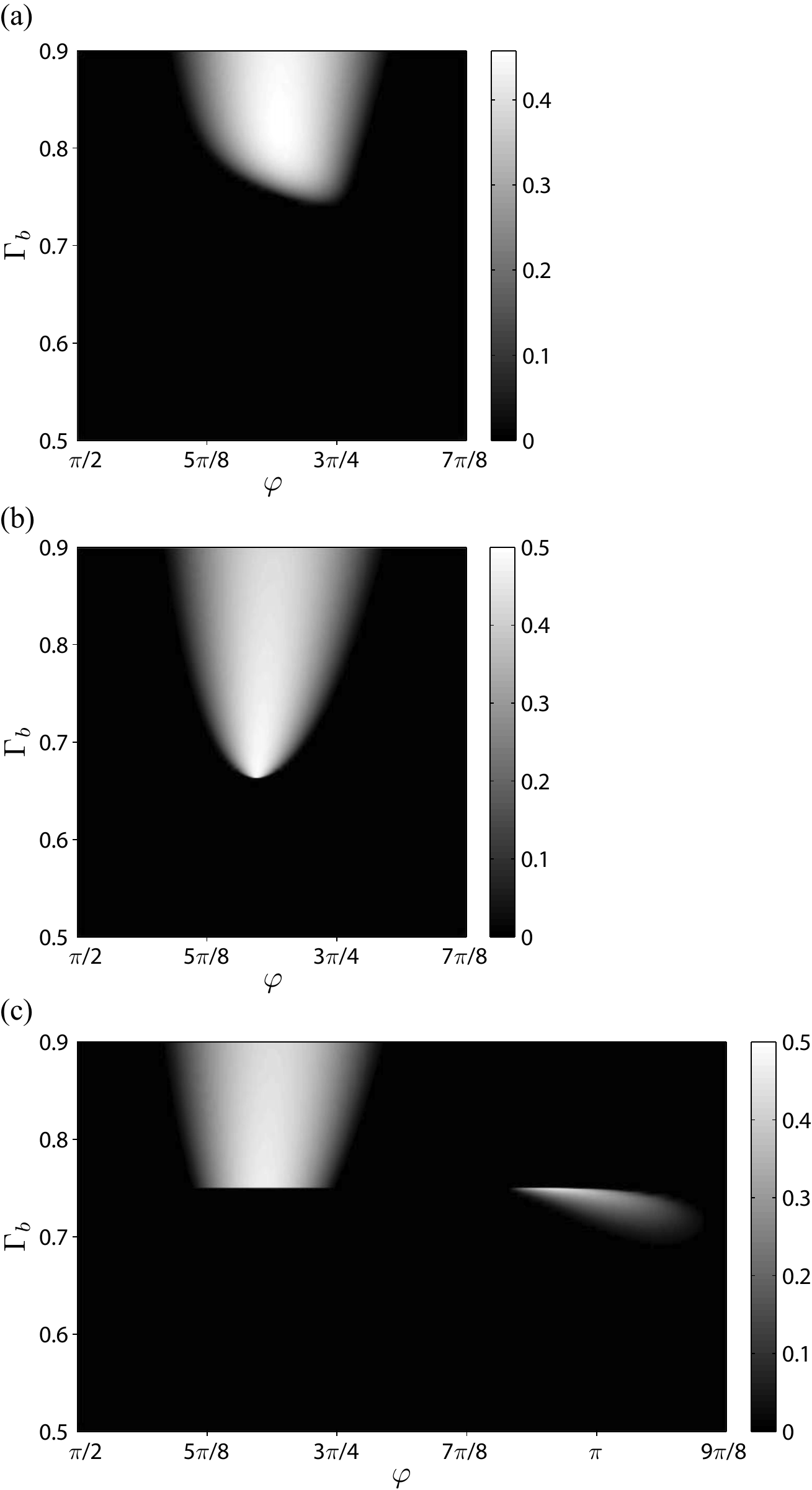}
\caption{(a) Entanglement measure $\mathrm{max}\{0,C_\varphi \}$ for $N=100$, $h=0.2$, $\lambda = 0.75$, and $\Gamma_b^\mathrm{c}=0.66$. (b) Entanglement measure $\mathrm{max}\{0,C_\varphi \}$ in the thermodynamic limit, with the same parameters, for the steady state (\ref{eq:diss_lmg_semicl_ss_sols}). (c) Entanglement measure $\mathrm{max}\{0,C_\varphi \}$ in the thermodynamic limit for the steady state (\ref{eq:diss_lmg_semicl_ss_sols_2}).} \label{fig:dissipative_cvarphi_first_order}
\end{figure}

We now consider again the entanglement measures $C_\varphi$ and $C_\mathrm{R}$, both numerically for finite $N$ and analytically for $N\gg1$, corresponding to the linearized regime. Figure~\ref{fig:dissipative_cvarphi_first_order}(a) shows a plot of $C_\varphi$ as a function of $\Gamma_b$ and $\varphi$ for $N=100$. We see that, well above the critical value $\Gamma_b=\lambda$, substantial entanglement is present over a broad range of angles $\varphi$. As $\Gamma_b$ approaches $\lambda$ from above, significant entanglement persists, but for a somewhat narrower range of angles $\varphi$. However, in the vicinity of $\Gamma_b=\lambda$ the entanglement diminishes rapidly for \emph{all} values of $\varphi$ as $\Gamma_b$ decreases further.

\begin{figure}[h!]
\centerline{\includegraphics[width=8.6cm]{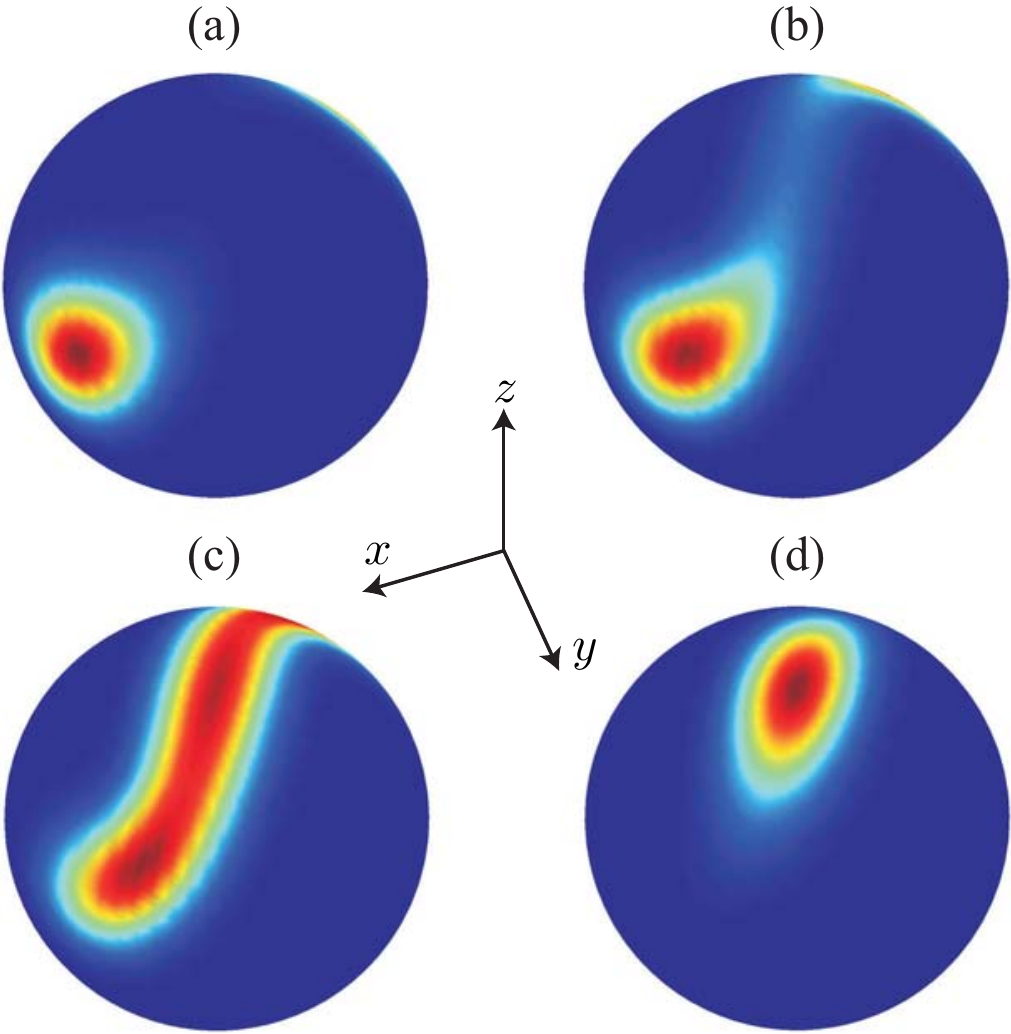}}
\caption{(Colour online) Steady state spin Q-function, $Q_\mathrm{s}(\eta)$, on the Bloch sphere for (a) $\Gamma_b = 0.55$, (b) $\Gamma_b = 0.65$, (c) $\Gamma_b = 0.705$, and (d) $\Gamma_b = 0.85$, with $N=50$, $h=0.2$, $\lambda = 0.75$ and $\Gamma_b^\mathrm{c}=0.66$. Note that dark blue corresponds to the minimum value of zero of $Q_{\rm s}(\eta)$ while dark red indicates the maximum value of $Q_{\rm s}(\eta)$.}\label{fig:dissipative_qspin_first_order}
\end{figure}

To help understand these results we again utilize the atomic coherent state representation and study the spin $Q$-function. In Fig.~\ref{fig:dissipative_qspin_first_order} we plot $Q_\mathrm{s}(\eta)$ on the Bloch sphere for a series of values of $\Gamma_b$ in the vicinity of the first-order transition. Well above the critical point $Q_\mathrm{s}(\eta)$ is a single-peaked function with little angular dependence. Correspondingly, $C_\varphi$ is non-zero over a broad range of $\varphi$, with a maximum close to $\varphi=3\pi/4$. Note that in contrast to the results found in \cite{Morrison07}, here there is a large shift of the optimum away from $\varphi=\pi/2$, since fluctuations in both the $y$ \emph{and} $x$ directions are significant (see Fig.~\ref{fig:dissipative_first_order_moments}).

As $\Gamma_b$ decreases towards the value $\lambda$, $Q_\mathrm{s}(\eta)$ becomes increasingly stretched along a direction between the $x$- and $y$-axes. As the critical value $\Gamma_b=\lambda$ is traversed, $Q_\mathrm{s}(\eta)$ changes from a single-peaked function to a triple-peaked function, corresponding to the existence of three stable steady states in the region $\Gamma_b^{\rm c}<\Gamma_b <\lambda$. As noted above, the range of $\varphi$ over which $C_\varphi$ remains finite narrows and then drops abruptly to zero as $\Gamma_b$ approaches $\lambda$ from above. This can be explained by noting that in the bistable region $\av{J_{\varphi}}=0$, and thus we again have $C_{\varphi} = 1-(4/N)\av{[\sin(\varphi)J_x + \cos(\varphi) J_y]^2}$. Since  both $\av{J_x^2}$ and $\av{J_y^2}$ are of order $j^2=N^2/4$ in this region (see Fig.~\ref{fig:dissipative_first_order_moments}), this severely restricts the range of $\Gamma_b$ and $\varphi$ for which $C_\varphi>0$.

As the dissipation strength is decreased further, eventually the critical point $\Gamma_b^{\rm c}$ is crossed and the central peak of $Q_\mathrm{s}(\eta)$ at the top of the Bloch sphere vanishes; we then recover the familiar two-lobed structure associated with the two semiclassical steady state amplitudes. The moments $\av{J_x^2}$ and $\av{J_y^2}$ are still of order $j^2=N^2/4$ in this region (and $\av{J_{\varphi}}=0$), and consequently $C_\varphi=0$ for all $\varphi$.

Now we briefly turn to the thermodynamic limit again where we can obtain analytic results for $N\gg1$ from the linearized HP model \cite{Morrison07}. Outside  the bistable region we can compute the entanglement by linearizing the fluctuations about the unique stable steady state. However, in the bistable region we can only compute the entanglement by linearizing about one or the other of the stable steady states.

In Fig.~\ref{fig:dissipative_cvarphi_first_order}(b) we display $C_\varphi$ as a function of the dissipation strength $\Gamma_b$ and the phase angle $\varphi$ for the choice of stable steady state given by (\ref{eq:diss_lmg_semicl_ss_sols}). We observe that, for $\Gamma_b> \Gamma_b^{\rm c}$, $C_\varphi$ is non-zero over a broad range of $\varphi$ centered around $\varphi = 3\pi/4$, whilst near the critical point $\Gamma_b^{\rm c}$ the corresponding range of $\varphi$ narrows. Outside the bistable region $C_\varphi$ is in reasonable agreement with the corresponding finite-$N$ results, however, as might be expected inside the bistable region, the results differ considerably, since linearization around simply one of the stable steady states is inadequate.

In Fig.~\ref{fig:dissipative_cvarphi_first_order}(c) we display $C_\varphi$ for the choice of stable steady state given by (\ref{eq:diss_lmg_semicl_ss_sols_2}). For $\Gamma_b> \lambda$, $C_\varphi$ is non-zero for a broad range of $\varphi$ centered around $\varphi = 3\pi/4$, which is also in reasonable agreement with the finite-$N$ results. However, in addition, in the bistable region $\Gamma_b^{\rm c}<\Gamma_b <\lambda$, we find that $C_\varphi$ is non-zero for a small range of $\varphi$. Specifically, we see that with decreasing $\Gamma_b$, away from the critical point, $C_\varphi$ becomes broader and its center moves toward $\varphi = \pi$. The existence of this second lobe and its behaviour can be described by considering the fluctuations in the linearized regime. In contrast to the finite-$N$ results one finds (not actually shown) that the fluctuations below the transition are less significant, thus giving $C_\varphi \neq 0$. The f
 act that the centre of $C_\varphi$ moves toward $\pi$ can be attributed to the dominating fluctuations in the $x$ direction for $\Gamma_b\ll\Gamma_b^\mathrm{c}$, as opposed to approximately equal fluctuations in the $x$ and $y$ direction for $\Gamma_b\simeq\Gamma_b^{\rm c}$.

\begin{figure}[h!]
\centerline{\includegraphics[width=6.5cm]{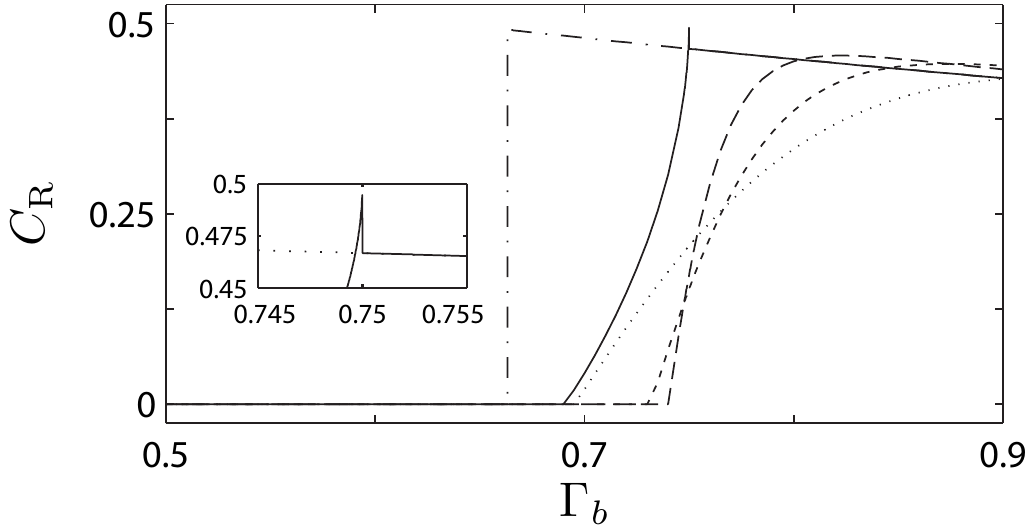}}
\caption{ Rescaled concurrence $C_\mathrm{R}$ for $N=25$ (dotted line), $N=50$ (short dashed line), $N=100$ (long dashed line), and in the thermodynamic limit [solid line for (\ref{eq:diss_lmg_semicl_ss_sols_2}), and dash dotted line for (\ref{eq:diss_lmg_semicl_ss_sols})], with $h=0.2$, $\lambda = 0.75$ and $\Gamma_b^\mathrm{c}=0.66$.} \label{fig:dissipative_concurr_first_order}
\end{figure}

Finally, in Fig.~\ref{fig:dissipative_concurr_first_order} we plot the rescaled concurrence $C_\mathrm{R}$ as a function of the dissipation strength $\Gamma_b$, for the system sizes of $N=25,50,100$ (set of dashed lines) and in the thermodynamic limit (for each choice of stable steady state in the bistable region). For finite $N$ the entanglement attains a peak value close to $\Gamma_b=\lambda$, while for $N\rightarrow\infty$ the entanglement peaks at either $\Gamma_b=\Gamma_b^{\rm c}$ or $\Gamma_b=\lambda$ for the different steady states, respectively. It is worth noting that the entanglement for the case corresponding to the more stable steady state, i.e., (\ref{eq:diss_lmg_semicl_ss_sols_2}), agrees more closely with the finite-$N$ result.

\section{Conclusion} \label{sect:conclusion}

We have shown the presence of quantum phase transitions occurring due to the variation of the strength of dissipation in two different collective spin systems. For the cooperative resonance fluorescence model we computed the steady state entanglement and showed how the results could be interpreted in terms of an atomic phase space distribution. In the dissipative LMG model we found that either a continuous or discontinuous phase transition occurs depending only on the ratio of the effective field and interaction strengths. The steady state entanglement was analyzed in detail and the modifications due to strong dissipation were interpreted with the help of the atomic phase space distribution. In the regime of the first-order phase transition we showed that bistable behaviour can occur as evidenced by the semiclassical analysis; specifically we showed that whilst a linearized analysis is generally inadequate in this regime, one of the two different stable solutions tends to dominate and this is also reflected by finite-$N$ calculations. Finally, we have also briefly explained how both of these models might be implemented using an ensemble of atoms that interacts with optical cavity and laser fields, and thus how the entanglement properties might be measured via the cavity output field.

In the future, it would be interesting to compare the phase-dependent entanglement measure considered here with the context-based entanglement measure studied in \cite{Carmichael04}, where entanglement is quantified by considering a continuous observation of the environment of a dissipative system. Specifically, the angle appearing in our phase-dependent measure can be associated with the phase of a local oscillator in the homodyne measurement considered in \cite{Carmichael04}. It would also be interesting to study entanglement criticality in dissipation-driven quantum phase transitions of other systems, such as generalized collective spin models \cite{EntLMGConcurrReview} subjected to strong dissipation, or in collective models with additional short-range interactions \cite{Lee04} and dissipation. Other indicative measures of the critical behaviour might also be interesting to consider such as the recently proposed mixed-state fidelity \cite{Ming08} or the operator fidelity susceptibility \cite{Zanardi07}. Another important future direction is the simulation of the full quantum model for much larger system size $N$, using, for example, quantum trajectory methods \cite{HowardsNewBook}, so that the approach towards the thermodynamic limit may be studied more carefully.

\begin{acknowledgements}
The authors thank H. Carmichael for discussions and acknowledge support from the Austrian Science Foundation and from the Marsden Fund of the Royal Society of New Zealand.
\end{acknowledgements}



\begin{thebibliography}{99}
\bibitem{Prefactor}
Note that we write the prefactor as $\gamma/N$ in anticipation of taking the thermodynamic limit, $N\rightarrow\infty$ at constant density.
\bibitem{Bistability}
Note that although for $\Gamma_b<\Gamma_b^{\rm c}$ there are already two stable steady states as given by Eq. (\ref{eq:diss_lmg_semicl_ss_sols_2}) this situation does not correspond to a bistability since the subsequent linearistion around each solution gives identical results.
\bibitem{Trajectories}
Outside the bistable region we find results very similar to those already obtained in Sec. \ref{sect:second_order} for the second-order phase transition.
\end{thebibliography}
\end{document}